\documentclass[journal,10pt]{IEEEtran}
\usepackage[T1]{fontenc}
\usepackage{cite}
\usepackage{flushend}
\usepackage{amsmath,graphicx}
\usepackage{amssymb}
\usepackage{algpseudocode}
\usepackage{amsfonts}
\usepackage{graphicx}
\usepackage{fancyhdr}  %
\usepackage{cases}
\usepackage{extarrows}
\usepackage{algorithm}
\usepackage{multirow,tabularx}
\usepackage{mathtools}
\usepackage{soul, color, xcolor}
\usepackage[english]{babel}
\usepackage{caption}
\usepackage{autobreak}
\usepackage{bm}

\newtheorem{theorem}{\bf Theorem}
\newtheorem{lemma}{\bf {Lemma}}
\newtheorem{proposition}{\bf {Proposition}}
\newtheorem{assumption}{\bf{Assumption}}
\newtheorem{definition}{\bf{Definition}}

\ifCLASSINFOpdf
\else
\fi
\begin{document}
\title{Coverage and Rate Analysis for Distributed RISs-Assisted mmWave Communications}
\author{Yuan Xu, Chongwen Huang,~\IEEEmembership{Member,~IEEE}, Wei Li, Yongxu Zhu,~\IEEEmembership{Senior Member,~IEEE}, Zhaohui Yang,\\ 
Jiguang He,~\IEEEmembership{Senior Member,~IEEE}, Jun Yang, Zhaoyang Zhang,~\IEEEmembership{Senior Member,~IEEE},\\ 
Chau Yuen,~\IEEEmembership{Fellow,~IEEE}, and Mérouane Debbah,~\IEEEmembership{Fellow,~IEEE}\vspace{-4mm}
\thanks{Y. Xu, C. Huang, Z. Yang and Z. Zhang are with the College of Information Science and Electronic Engineering, Zhejiang University, Hangzhou 310027, China, and with Zhejiang Provincial Key Laboratory of Info. Proc., Commun. \& Netw. (IPCAN), Hangzhou 310027, China (E-mails: \{yuan\_xu, chongwenhuang, zhaohui\_yang, ning\_ming\}@zju.edu.cn).\\
\indent L. Wei and C. Yuen are with the School of Electrical and Electronics Engineering, Nanyang Technological University, Singapore 639798 (E-mails: l\_wei@ntu.edu.sg, chau.yuen@ntu.edu.sg).\\
\indent Y. Zhu is with the National Communications Research Laboratory, Southeast University, Nanjing 210096, China (E-mail: yongxu.zhu@seu.edu.cn).\\
\indent J. He is with the Technology Innovation Institute, 9639 Masdar City, Abu Dhabi, UAE (E-mail: jiguang.he@tii.ae).\\
\indent J. Yang is with the State Key Laboratory of Mobile Network and Mobile Multimedia Technology, Shenzhen, 518055, China. J. Yang is also with Wireless Product R\&D Institute, ZTE Corporation, Shenzhen 518057, China (E-mail: yang.jun10@zte.com.cn).\\
\indent M. Debbah is with the Department of Electrical Engineering and Computer Science and the KU 6G Center, Khalifa University, Abu Dhabi 127788, UAE, and also
with CentraleSupelec, University Paris-Saclay, 91192 Gif-sur-Yvette, France (E-mail: Merouane.Debbah@ku.ac.ae).
}}

\maketitle                                                                  
\thispagestyle{empty}
\pagestyle{empty}
\begin{abstract}
The millimeter wave (mmWave) has received considerable interest due to its expansive bandwidth and high frequency. However, a noteworthy challenge arises from its vulnerability to blockages, leading to reduced coverage and achievable rates. To address these limitations, a potential solution is to deploy distributed reconfigurable intelligent surfaces (RISs), which comprise many low-cost and passively reflected elements, and can facilitate the establishment of extra communication links. In this paper, we leverage stochastic geometry to investigate the ergodic coverage probability and the achievable rate in both distributed RISs-assisted single-cell and multi-cell mmWave wireless communication systems. Specifically, we first establish the system model considering the stochastically distributed blockages, RISs and users by the Poisson point process. Then we give the association criterion and derive the association probabilities, the distance distributions, and the conditional coverage probabilities for two cases of associations between base stations and users without or with RISs. Finally, we use Campbell's theorem and the total probability theorem to obtain the closed-form expressions of the ergodic coverage probability and the achievable rate. Simulation results verify the effectiveness of our analysis method, and demonstrate that by deploying distributed RISs, the ergodic coverage probability is significantly improved by approximately 50\%, and the achievable rate is increased by more than 1.5 times.
\end{abstract}
\vspace{-2mm}	


\vspace{-3mm}
\section{Introduction}\label{sec:intro}
\vspace{4.4mm}
The millimeter wave (mmWave) technology has emerged as a pivotal technology for the next-generation mobile communication networks, attributed to its abundant spectrum resources and reduced spectrum interference compared to the microwave band\cite{Jameel8594703,Yang8613274,Lodro8673447}. Specifically, the high-frequency nature indicates a large bandwidth, which improves the communication throughput to a considerable extent\cite{Ghosh6824746}. Besides, high frequencies result in short wavelengths, which allow for the use of smaller antennas, making it feasible to deploy large antenna arrays in a relatively small physical space. It is particularly beneficial to multiple-input multiple-output (MIMO) communication systems\cite{9779586,9136592,8523816}.
\par
However, short wavelengths make mmWave signals highly susceptible to atmospheric absorption, rain attenuation, obstruction, etc., leading to substantial penetration loss and path loss\cite{9902991,Niu07228,Wei7000981}. A feasible solution to address these challenges and ensure reliable communications is deploying distributed reconfigurable intelligent surfaces (RISs)\cite{Huang8741198,9899454}. Specifically, a RIS is a planar antenna array comprising numerous low-cost and passive reflective elements. Each unit or element, under the control of an intelligent controller, can induce tunable amplitude and phase shift to reflect incident electromagnetic waves. Consequently, RISs have the capability to reshape the wireless propagation environment. Further, they operate in an energy-efficient manner since RISs only passively change the directions of the received signals. This results in a significant reduction in energy consumption, which is orders of magnitude lower than conventional active antenna arrays\cite{Jung9357969,9676676}. Moreover, the distributed RISs provide more transmission links compared to a centralized RIS, effectively guaranteeing the quality of service for users\cite{9762646}. Therefore, we are interested in how much gain the distributed RISs can bring to mmWave communications.
\par 
Nonetheless, analyzing the communication performance in a distributed RISs-assisted mmWave system remains a challenge. Recently, stochastic geometry has become a focal point of research on distributed systems due to its ability to capture the inherent spatial randomness in diverse types of wireless networks and naturally extend the model accounting for other uncertainties, such as fading, shadowing, and power control\cite{Haenggi5226957}. Moreover, stochastic geometry has been proven to be a valuable tool by providing a unified mathematical framework to describe the behavior of wireless communication systems\cite{Win2006ErrorPO,Win2009AMT,Haenggi5226957}, and can even yield closed-form expressions in certain scenarios\cite{Di7105406,Zhang7377022}.
\par 
Based on the above discussions, this paper leverages stochastic geometry to investigate the potential gain of the ergodic coverage probability and the achievable rate in the distributed RISs-assisted single-cell and multi-cell mmWave communication systems. In the following subsections, we first review the relevant research works and then summarize our contributions.
\vspace{-3mm}
\subsection{Related Works}
\vspace{1.5mm}
There has been some literature focusing on the analysis of RISs-assisted communication systems\cite{Di9140329,5d9a821,Tang9206044,Karasik2019,Perovi2019,Zhang9110912,Huang2018,Jung9357969,Hou9171881}. Specifically, the authors of \cite{Di9140329,5d9a821,Tang9206044} modeled the path loss for RIS reflective links in free-space scenarios, omitting considerations of scattering, shadowing, and reflections. The study in \cite{Karasik2019} derived the capacity of RISs-assisted single-input multiple-output communication systems. It demonstrated that a joint information coding scheme involving transmitted signals and RIS configurations is necessary to achieve channel capacity. The authors of \cite{Perovi2019} investigated RISs-assisted indoor mmWave communications, proposing two schemes to maximize channel capacity. In \cite{Zhang9110912}, fundamental capacity constraints of RISs-assisted point-to-point MIMO communication systems were described. In \cite{Huang2018}, Huang \textit{et al.} maximized the sum rate in RISs-enhanced multi-user multiple-input single-output downlink communications, while authors in \cite{Jung9357969} analyzed the asymptotic optimality of the achievable rate in the RISs-assisted downlink system with practical constraints, and proposed a modulation scheme for RISs that does not interfere with existing users. Furthermore, a new resource allocation algorithm was designed, considering user scheduling and power control. \cite{Hou9171881} employed signal cancellation to design RIS beamforming in MIMO non-orthogonal multiple access networks. To evaluate the performance of the network, \cite{Hou9171881} derived closed-form expressions of the outage probability and ergodic rate, and also evaluated the network performance with finite resolution design of RIS phase shifts and amplitude coefficients.
\par 
Generally, stochastic geometry has been used to analyze mmWave communication schemes\cite{bai6840343,Bai6932503,Rebato8628991}. To be more specific, the receiver and transmitter locations were stochastically modeled using the Poisson point process (PPP) in \cite{bai6840343}. By considering typical receivers at the origin, the authors simplified the analysis of the average performance of the network, provided bounds on network performance metrics, which approached realistic scenarios. In \cite{Bai6932503}, the authors utilized the line-of-sight (LoS) probability function proposed in \cite{bai6840343} to model the locations of LoS and non-line-of-sight (NLoS) base stations (BSs) as two independent inhomogeneous PPPs. The dense network case was further analyzed by applying a simplified system model, where the LoS area of a user was approximated as a fixed LoS sphere. In \cite{Rebato8628991}, the authors proposed an analytical framework that included realistic channel models and antenna element radiation patterns. Then, gain models were established on the desired signal beam and the interfering signal beam as well, respectively.
\par 
Further, there are also a few works on the analysis of RISs-assisted communication scenes using stochastic geometry\cite{Kishk2001,Boulogeorgos9320587,Nemati9224676,Zhu9110835}. In \cite{Kishk2001}, the authors neglected the small-scale channel fading model, deployed RISs as coatings on a portion of blockages, and leveraged stochastic geometry to analyze the impact of coating blockages with RISs on the coverage probability of cellular networks. In \cite{Boulogeorgos9320587}, the authors use system models considering terahertz links and RIS characteristics, as well as a new generic end-to-end channel fading formula to analyze its performance. Furthermore, the authors in \cite{Nemati9224676} studied the coverage probability of RISs-assisted large-scale mmWave cellular networks leveraging stochastic geometry and derived the closed forms of the peak received power expression of RIS and the downlink signal-to-interference ratio coverage expression, while stochastic geometry was employed for the performance analysis of realistic two-step user associations over
mmWave networks in \cite{Zhu9110835}. Doubtlessly, these above-mentioned works consistently verified that RISs can theoretically provide unprecedented improvements in both the achievable rate and energy efficiency.
\vspace{-4mm}
\subsection{Contributions}
\vspace{2.mm}
The main contributions of this paper are the analysis and validation of the gains in the ergodic coverage probability and the achievable rate for the distributed RISs-assisted single-cell and multi-cell mmWave communications, by employing the stochastic geometry theory. The findings of this study offer valuable insights and guidance for the practical deployment of RISs in mmWave communication networks. To the best of our knowledge, this is the comprehensive work to systematically investigate the impact of distributed RISs on both single-cell and multi-cell system performance. 
\par 
In more detail, we start with scene modeling. Ignoring the height information, we model the blockages using the line Boolean model. The central points distribution of the RISs, blockages, and users are all modeled as PPPs. For the BS model, the BS in the single-cell scenario is defined as centered in the two-dimensional (2D) Euclidean plane, whereas in the multi-cell scenario, the distribution of multiple BSs is defined as a PPP. Then we establish the system model of the networks, employing the statistical model of large-scale channel fading and the segmented approximate exponential model of small-scale fading proposed in \cite{Rebato8628991}.
\par 
Subsequently, we propose the user-BS association criterion and separately analyze two cases: users associating with the serving BS directly or through RIS reflection. In the single-cell scenario, we thin the RIS distribution model into several inhomogeneous PPPs using the LoS probability and derive the association probabilities, the distance distributions, and the conditional coverage probabilities. Using the total probability theorem and Campbell's theorem, we combine the two cases to derive the ergodic coverage probability. In the multi-cell scenario, we focus on a typical user for simplicity since each user shares the same statistical characteristics. Similar to the single cell, we analyze the conditional coverage probabilities for the two association cases respectively and derive the closed-form multi-cell ergodic coverage probability. Moreover, we derive closed-form expressions for the achievable rate by transformations between the expectation and distribution functions.
\par 
Simulation results validate that the deployment of distributed RISs reduces the mmWave system blind area ratio by over 70\%, improves the ergodic coverage probability by about 50\%, and achievable rate by more than 1.5 times. Notably, we comprehensively consider realistic small-scale fading, large-scale fading channel gain models, and multi-BS interference as well.


\begin{figure}[t]
	\begin{center}
			\centerline{\includegraphics[width=0.49\textwidth]{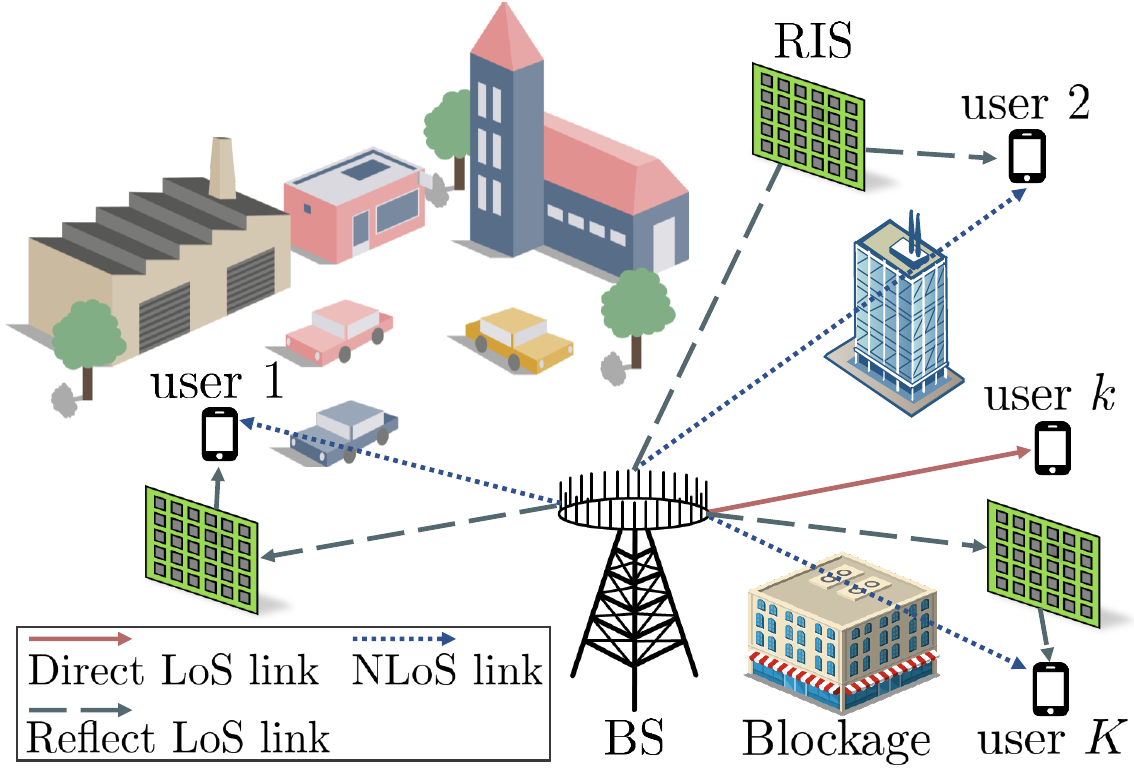}}  \vspace{-1mm}
			\caption{Downlink distributed RISs-assisted mmWave single-cell communication scenario.  }
			\label{fig:scene_s} \vspace{-8mm}
		\end{center}
\end{figure} 
\vspace{-5mm}
\section{System Model}\label{sec:format}
\vspace{3.5mm}
In this section, we introduce the scene and system model for downlink distributed RISs-assisted single-cell and multi-cell mmWave wireless communication systems. Disregarding the height information, we introduce the concept of the cell virtual radius for the convenience of explanation, i.e., the equivalent circle radius $r_v$ of the average cell size. 
\par
Firstly, we define the distribution of the blockages, users, and RISs. Blockages, especially buildings in urban areas, often have random shapes. To capture the spatial randomness and shape irregularity, they are commonly modeled as random object processes (ROPs) in the field of random shape theory\cite{Richard10}. However, the analysis of ROPs can be intractable, particularly when there are correlations between objects or when the shape, position, and orientation of the objects are interrelated. Thus, we model the blockages using the line Boolean model\cite{bai6840343} in the following assumption.
\begin{assumption}
\textit{(Blockage Process)} In the line Boolean model, each blockage is represented by a line segment. The central positions of all line segments are modeled by a PPP $\Phi_{b}=\{z_i\}\in\mathbb{R}^2 $ with a density of $\lambda_{b} $. Here, $z_i$ represents the position coordinate of the center of the $i$-th blockage, and $\Phi_b $ represents the set of all blockage center coordinates. The length of each line segment, denoted as $L_i$, follows a uniform distribution $\mathbf{U}[L_{min}, L_{max}] $ with a mean value of $\mathbb{E}[L] $, where $L_{min} $ and $L_{max} $ represent the minimum and maximum lengths of all line segments, respectively. Furthermore, the orientation angle of each blockage, defined as the angle between the line segment and the positive direction of the $x$-axis, follows a uniform distribution $\phi_{b,i}\sim\mathbf{U}[0,2\pi] $.
\end{assumption}
\par
The users can be considered as static in a short time interval, and their positions are modeled by a PPP expressed as $\Phi_{u}=\{u_i\}\in\mathbb{R}^2$, with a density of $\lambda_{u}(\xi)$, and each user is equipped with $N_u$ antennas. Here, $u_i$, $\Phi_{u}$, and $\xi$ represent the position coordinate of the $i$-th user, the set containing the location coordinates of all users, and the distance variable between the BS and users, respectively. The positions of RISs are also modeled using a PPP denoted as $\Phi_{R}=\{R_i\}\in\mathbb{R}^2$, with a density of $\lambda_{R} $, where $R_i$, $\Phi_{R}$, and $N_R$ represent the position of the $i$-th RIS, the set of all RIS coordinate points, and the number of elements of each RIS, respectively. It is assumed that RISs will assist communications only when the direct links are NLoS, and can both reflect and transmit signals, providing 360-degree coverage.
\par
In the single-cell scenario, the BS is positioned at the center of the area and is equipped with $N_{BS}$ antennas. In the multi-cell scenario, the coordinates of the BSs follow a PPP $\Phi_{Y}=\{Y_i\}\in\mathbb{R}^2$, with a density of $\lambda_{Y}$, where $Y_i$, $\Phi_{Y}$, and $N_{BS}$ respectively denote the position of the $i$-th BS, the set of all BS coordinates, and the number of antennas for each BS.
\par
For the signal model, we first introduce the large-scale and small-scale fading models.
\begin{assumption}
	\textit{(Large-Scale Fading Channel Gain)} The LoS channel gain for large-scale fading follows the 3GPP model as
	\begin{equation}
		g=10^\alpha d^{-\beta},  
	\end{equation}
	where $\alpha=-2.8-2\log_{10}(\frac{f_{c}}{1e9})$, and $f_{c}$ denotes the working frequency. The parameter $\beta$ represents the exponent of path loss, while $d$ represents the distance between the transmitter and the receiver.
\end{assumption}
\begin{assumption}
	\textit{(Small-Scale Fading Channel Gain)}\label{smallscale} The small-scale fading channel gain is assumed to follow an exponential distribution \cite[Remark 5]{Rebato8628991}, with a mean value of $\rho_{t,r}$, where $\rho_{t,r}$ is the directional gain approximated by the segmented antenna model, which is given by
	\begin{equation}
		\rho_{t,r}=\left\{
		\begin{aligned}
			&M_tM_r,\ \text{w.p.}\  \frac{\psi_t}{2\pi}\frac{\psi_r}{2\pi}\\
			&M_tm_r,\ \text{w.p.}\  \frac{\psi_t}{2\pi}(1-\frac{\psi_r}{2\pi})\\
			&m_tM_r,\ \text{w.p.}\  (1-\frac{\psi_t}{2\pi})\frac{\psi_r}{2\pi}\\
			&m_tm_r,\ \text{w.p.}\  (1-\frac{\psi_t}{2\pi})(1-\frac{\psi_r}{2\pi}).
		\end{aligned}
		\right.
	\end{equation}
 The subscripts $t$ and $r$ denote the transmitter and receiver, respectively. $M_t$ and $M_r$ denote their main lobe directivity gains, while $\psi_t$ and $\psi_r$ denote their corresponding main lobe beam widths. We consider the isotropic antennas, the main lobe directivity gains are equal to the number of antenna elements, i.e., $M_t=N_t$, $M_r=N_r$. Meanwhile, the side lobe gains are expressed as $m_t=1/\sin^2(\frac{3\pi}{2\sqrt{N_t}}) $ and $m_r=1/\sin^2(\frac{3\pi}{2\sqrt{N_r}}) $, and the angle of arrival (AOA) and angle of departure (AOD) follow a uniform distribution on $(0,2\pi)$.
\end{assumption}
\par
In addition, we assume that the data transmission link between the BS and the user has already been implemented using existing beamforming techniques like zero forcing\cite{Spencer1261332}, minimum mean square error\cite{Nguyen8048606}, single value decomposition\cite{Li7996970}, etc., which also enable us to effectively eliminate interference between multiple users within the same cell. In other words, the interference term can be disregarded within the single cell. In a multi-cell scenario, the interfering signals only come from the non-serving BSs through links without beam alignment. In the following, we provide the signal model in single-cell and multi-cell scenarios.
\vspace{-2.mm}
\subsection{Single-Cell Scene}\label{subsec:s-Scene model} 
\vspace{1.7mm}
We consider a downlink distributed RISs-assisted mmWave communication cell with a virtual radius of $r_v=R$, as illustrated in Fig.~\ref{fig:scene_s}. A single BS serves $K$ users with the assistance of $I$ RISs. It is assumed that the BS possesses the location information of the RISs, and there is always a LoS link between the BS and the RISs. Next, we define the signal-to-interference-plus-noise ratio (SINR) of the direct links and the RIS reflection links. 

\par
For a user located at a distance $\xi$ from the BS, the large-scale fading channel gain of the direct link is denoted as $g^s_d=\frac{10^\alpha}{\xi^\beta}$. The received signal at the user, denoted as $y^s_d$, has a power $|y^s_d|^2=g_d^s P_0 h^s_d$. Here, $P_0$ denotes the BS transmit power to each user, and $h^s_d$ denotes the small-scale fading channel gain of the BS-user link, which follows an exponential distribution with a mean of $N_{BS}N_u$, denoted as $h^s_d\sim \mathbf{Exp}(N_{BS}N_u)$. Since the interference from other users has been eliminated, the SINR is simplified to the signal-to-noise ratio (SNR). Therefore, the SINR for the direct link is expressed as 
\begin{equation}
    \gamma^s_d=\frac{10^\alpha\cdot P_0\cdot h^s_d}{\xi^\beta\sigma^2},
\end{equation}
where $\sigma^2$ denotes the noise power.
\par
In the case where the BS reflects the signal to the user through the $l$-th RIS $R_l$, we define the link as a reflective LoS link. The large-scale fading channel gain and the received signal power of the reflective LoS link are expressed as $g^s_{I_l}=\frac{10^{2\alpha}}{(\dot{s}_l\dot{r}_l)^\beta}$, and $|y^s_{I_l}|^2=\frac{10^{2\alpha}P_0 h_{\dot{s}_l} h_{\dot{r}_l}}{(\dot{s}_l\dot{r}_l)^\beta} $, respectively. Here, $\dot{s}_l$ and $\dot{r}_l$ denote the distance of the BS-$R_l$ link and the $R_l$-user link, while $h_{\dot{s}_l}\sim \mathbf{Exp}(N_{BS}N_R)$ and $h_{\dot{r}_l}\sim \mathbf{Exp}(N_RN_u)$ denote the small-scale fading channel gains of the BS-$R_l$ and $R_l$-user links respectively. The SINR of the BS-$R_l$-user link is expressed as
\begin{equation}
    \gamma^s_{I_l}=\frac{10^{2\alpha}P_0 h_{\dot{s}_l} h_{\dot{r}_l}}{(\dot{s}_l\dot{r}_l)^\beta\sigma^2}.
\end{equation}
\vspace{-6.7mm}
\subsection{Multi-Cell Scene}
\vspace{3mm}
Different from the single-cell scenario, we consider an infinitely large 2D downlink distributed RISs-assisted mmWave communication network having multiple cells, as illustrated in Fig.~\ref{fig:P_R_s}. Since each BS in the network serves one cell and the BS density is $\lambda_Y$, the average size of each cell is expressed as $1/\lambda_Y$\cite{Baccelli8187456}, thus the virtual radius of each cell is $r_v=\sqrt{\frac{1}{\lambda_Y\pi}}$. The reason for introducing $r_v$ is that it provides an equivalent description of the BS density in the network. That is, a larger $r_v$ indicates a lower BS density. Moreover, the virtual cell radius is directly related to the BS spacing employed in practical BS planning\cite{Bai6932503}.
\par 
For the sake of theoretical analysis, we consider a typical user at the origin of the coordinate system, since the locations of different users are independent and relatively stationary over short time intervals, thus the statistical performance of any user is considered consistent with that of users at other locations. It is worth noting that the results of this analysis are applicable to different point distributions in which only users are independent. Firstly, we divide the BS PPP $\Phi_Y$ into three parts based on location-dependent thinning and the LoS probability.
\begin{proposition}\label{thinning}
	\textit{(Location-Dependent Thinning}\cite{RePEc}\textit{)} For a homogeneous PPP $\Phi$ with density $\lambda$, thinning with probability $g(x)$, where $g(x)$ only depends on the point $x$ and not relevant to the remaining points, we can obtain the inhomogeneous PPP $\tilde{\Phi}$ with density $\tilde{\lambda}=[1-g(x)]\lambda$, while the removed points constitute an inhomogeneous PPP with density $\overline{\lambda}=g(x)\lambda$.
\end{proposition}
\begin{assumption}
	\textit{(LoS Probability) \label{PLoS}}For each link, the LoS probability is only related to the distance and the blockage information, but not to the channel state information. Besides, it is assumed that the influence of the blockage correlation between different links can be ignored. Let $Z$ be the number of blockages intersecting the link, according to the line Boolean model of blockages, we have $\mathbb{E}[Z]=\frac{2\lambda_b \mathbb{E}[L]d}{\pi}$\cite{bai6840343}. Consequently, the probability of existing a LoS link at a distance $d$ is equivalent to the probability of no blockage intersecting that link, which is the void probability when $Z=0$, denoted as\vspace{-1mm}
 \begin{equation}
     P_{LoS}(d)=\exp(-\mathbb{E}[Z])=\exp(-\frac{2\lambda_b \mathbb{E}[L]d}{\pi}).
 \end{equation}
\end{assumption}
\par
\begin{lemma}
	\textit{(RIS Assist Probability)} In the scenario depicted in Fig.~\ref{fig:P_R_s}, the distances of the BS-user, BS-RIS, and RIS-user links are denoted as $\xi$, $s$, and $r$, respectively. Since there is always a LoS link between each BS and each RIS, the probability that the RIS can provide a reflection link for the BS and the user is $P_{LoS}(r)$. 
\end{lemma}
\par
According to \textbf{Proposition \ref{thinning}}, we obtain the inhomogeneous PPP $\Phi_R^L$ by preserving the RIS PPP $\Phi_R$ with the probability $P_{LoS}(r)$, which represents the subset of RISs that assist the communication between the BS and the user. The density of $\Phi_R^L$ is expressed as $\lambda_R^L=\lambda_R\cdot P_{LoS}(r) $. 
\begin{lemma} \textit{(Multi-Cell Reflection Probability)}
	The probability that this typical user can communicate with the BSs through RISs reflection is\vspace{-2mm}
	\begin{equation}
		P_R^m=1-\exp(-\int_0^\infty \lambda_R^L 2\pi r\mathrm{d}r).
	\end{equation}
 That is, there is at least one reflective LoS link, the probability of which is equal to 1 minus the probability that $\Phi_R^L$ is an empty set.
\end{lemma}
\par
Thus, we can perform location-dependent thinning on the BS PPP $\Phi_Y$ with the probability that the direct link is LoS $P_{LoS}(\xi)$ and the multi-cell reflection probability $P_R^m$, where $\xi$ is the distance between the typical user and BS in $\Phi_Y$. Then three inhomogeneous PPP can be obtained, which are LoS BSs $\Phi_Y^L$, NLoS BSs with reflective LoS links $\tilde{\Phi}_Y^N$, and idle BSs which cannot communicate with the typical user $\Phi_Y^{N_I}=\Phi_Y-\Phi_Y^L-\tilde{\Phi}_Y^N$. Correspondingly, the densities are $\lambda_Y^L=\lambda_YP_{LoS}(\xi) $, $\tilde{\lambda}_Y^N=\lambda_Y^NP_R^m=\lambda_Y(1-P_{LoS}(\xi))P_R^m$ and $\lambda_Y^{N_I}=\lambda_Y(1-P_{LoS}(\xi))(1-P_R^m) $.
\par 
For the typical user, the signal power received from the LoS BS $Y_i\in\Phi_Y^L$ through the direct link is expressed as\vspace{-1mm}
\begin{equation}
    |y_{d_{i}}|^2=g_{d_{i}}P_0 h_{d_i},
\end{equation}
where $\xi_i$ denotes the distance from the BS $Y_i$ to the typical user, while $g_{d_{i}}=\frac{10^\alpha}{\xi_i^\beta}$ and $h_{d_i}\sim \mathbf{Exp}(\rho_{BS,u})$ denotes the large- and small-scale fading channel gain of the direct link from the BS $Y_i$ to the user. The signal power received from the NLoS BS $Y_j\in\tilde{\Phi}_Y^N$ through the $l$-th RIS reflection link is expressed as \vspace{-2mm}
\begin{equation}
    |y_{I^l_{j}}|^2=g_{I^l_{j}}P_0 h_{s_{j,l}}h_{r_{l}},
\end{equation}
where $s_{j,l} $ and $r_{l} $ represent the BS $Y_j$-RIS $R_l$ and $R_l$-user distance. Moreover, $g_{I^l_{j}}=\frac{10^{2\alpha}}{(s_{j,l}r_{l})^\beta}$ represents the large-scale fading channel gain of the BS $Y_j$-RIS $R_{l}$-user reflection link, while $h_{s_{j,l}}\sim \mathbf{Exp}(\rho_{BS,R})$ and $h_{r_{l}}\sim \mathbf{Exp}({\rho_{R,u}})$ represent the small-scale fading channel gain of the $Y_j$-$R_{l}$ and $R_{l}$-user link in the BS $Y_j$-RIS $R_{l}$-user reflection link.
\par 
Next, we define the association criteria between the user and the serving BS. It is assumed that the user density is significantly higher than the BS density, and a BS can be associated with multiple users but serves one user at a time. The association links can be either direct or reflective LoS links, which provide the maximum channel gain.
\begin{definition}
	(Association Criterion) \label{asscocition criterion}Each user associated with the BS providing the maximum signal power (either in $\Phi_Y^L$ or in $\tilde{\Phi}_Y^N$), which can be expressed as
	\begin{equation}
		|y^*|^2=\max(\max\limits_{Y_i\in\Phi_Y}|y_{d_{i}}|^2,\max\limits_{{Y_j\in\tilde{\Phi}_Y^N}}(\max\limits_{R_l\in\Phi_R^L} |y_{I^l_{j}}|^2)).
	\end{equation}
	In order to establish the feasible analysis equations, we made some appropriate simplifications by replacing the small-scale fading channel gain with its time-averaged value. Therefore, the association criterion can be simplified as 
 \begin{equation}
     \begin{split}
         |y^*|^2\!=\!N_{BS}N_u\max(\max\limits_{Y_i\in\Phi_Y}g_{d_{i}},\!\max\limits_{Y_j\in\tilde{\Phi}_Y^N} g_{I_j}N_R^2) ,
     \end{split}
 \end{equation}
where $g_{I_j}=\max\limits_{R_l\in\Phi_R^L} g_{I^l_{j}}$.
\end{definition}
\par
We define the serving BS of the typical user as $Y_0$ and assume that the link between the user and BS $Y_0$ has been beam-aligned, while non-serving BSs interfere with the user through misaligned links, i.e., the AOA and AOD are random and satisfy the uniform distribution on $[0,2\pi]$.
\par
When the user communicates with the serving BS $Y_a\in\Phi_Y$ on a direct LoS link, the SINR is defined as
\begin{equation}
	\gamma_d=\frac{g_{d_a}P_0 h_{d_a}}{\sigma^2+\sum\limits_{Y_i\in\Phi_Y^L\setminus Y_a} {g_{d_i}P_0 h_{d_i}}+\sum\limits_{Y_j\in\tilde{\Phi}_Y^N} {g_{I_j}P_0 h_{s_{j}}h_{r_{j}}} },
\end{equation}
where $h_{s_{j}}=h_{s_{j,l_j}}$, and $l_j=\arg\max\limits_{R_l\in\Phi_R^L} g_{I^l_{j}}$, indicating that we only consider interference from the NLoS BS through the strongest RIS reflection link. Meanwhile, $\Phi_Y^L\setminus Y_a$ 
denotes the PPP that removes BS $Y_a$ from the BS PPP 
$\Phi_Y^L$.
\par
In the case of the user communicating with the serving BS $Y_a$ on a reflective LoS link, the SINR is defined as
\begin{equation}
\gamma_I=\frac{g_{I_a}P_0 h_{s_{a}}h_{r_{a}}} {\sigma^2+\!\sum\limits_{Y_i\in\Phi_Y^L} {g_{d_i}P_0 h_{d_i}}+\!\sum\limits_{Y_j\in\tilde{\Phi}_Y^N\setminus Y_a} {g_{I_j}P_0 h_{s_{j}}h_{r_{j}}} },
\end{equation}
where $\tilde{\Phi}_Y^N\setminus Y_a$ 
denotes the PPP that removes BS $Y_a$ from the BS PPP 
$\tilde{\Phi}_Y^N$.
\vspace{-3mm}
\section{Performance Analysis}
\vspace{3.5mm}
Here, we analyze communication performance in the distributed RISs-assisted mmWave wireless cellular system. Firstly, we introduce the intrinsic properties of the PPP, namely the void probability and the contact distribution.
\begin{proposition}\label{voidcontact}
	\textit{(Void Probability and Contact Distribution}\cite{MOLTCHANOV20121146}\textit{)} 
	For any inhomogeneous PPP $\Phi_{in}\in\mathbb{R}^2$ with density $\lambda_{in}(x)$, the number of points on any given region $B\subset\mathbb{R}^2$ is a random variable satisfying the Poisson distribution with mean $\int_B \lambda_{in}(x) \mathrm{d}x$, and its void probability is\vspace{-2mm}
	\begin{equation}
		P_{void}=\exp(-\int_B \lambda_{in}(x) \mathrm{d}x).
	\end{equation}
	Its contact distribution is\vspace{-2mm}
	\begin{equation}
		F_{R_c}=1-\exp\left(-\int_B \lambda_{in}(x) \mathrm{d}x\right).
	\end{equation}
\end{proposition}
\vspace{-4.7mm}
\subsection{Single-Cell Performance}
\vspace{3mm}
\begin{figure}
    \begin{center}
        \centerline{\includegraphics[width=0.5\textwidth]{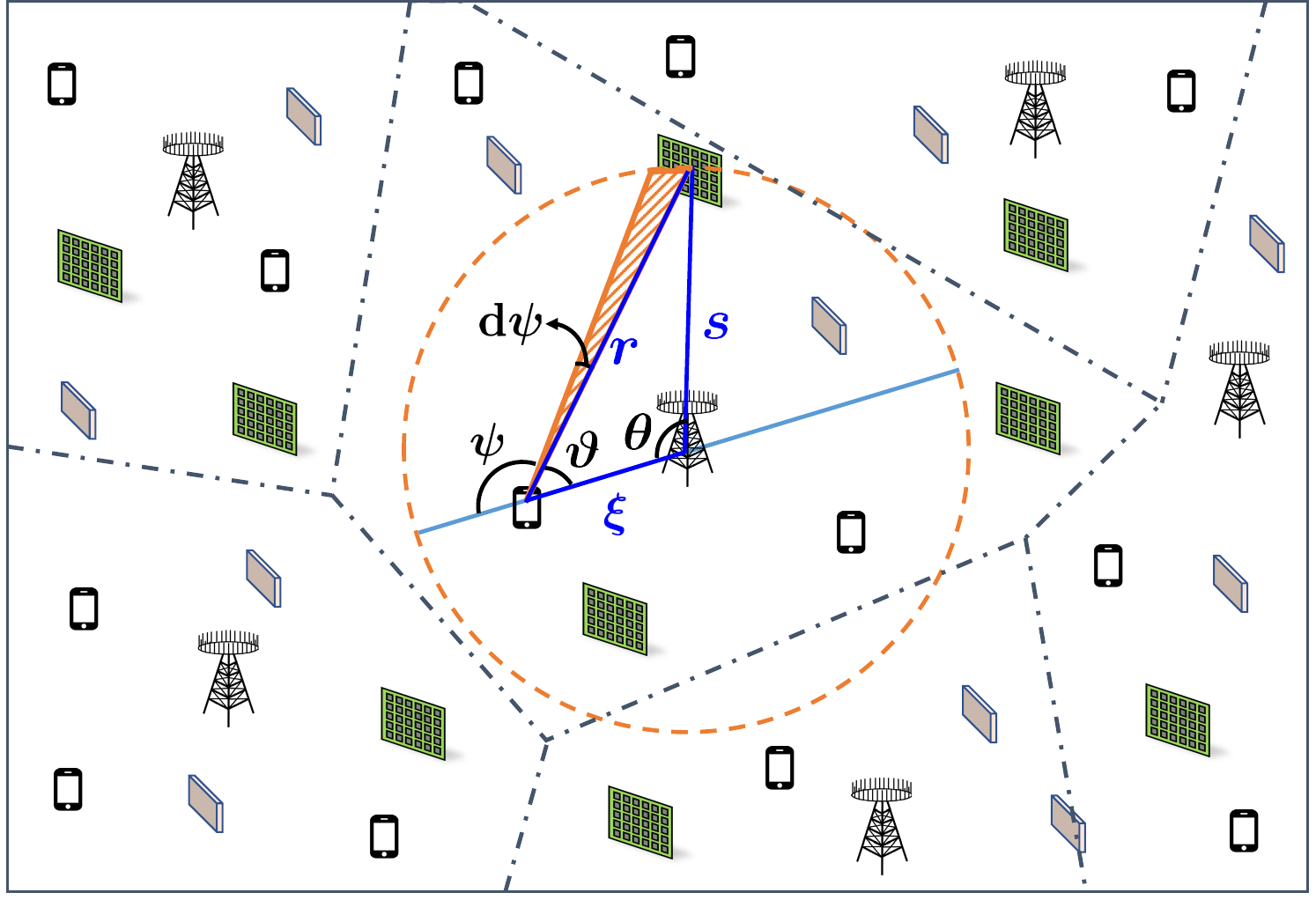}}  \vspace{-0mm}
        \caption{The spatial location between an arbitrary BS, RIS, and user.}
        \label{fig:P_R_s} 
    \end{center}\vspace{-4mm}
\end{figure}
We study two cases of the direct association and RIS reflection association. Since the multiplicative path loss of the RIS reflection link is several orders of magnitude larger than that of the direct link, the association criterion degenerates. When a direct LoS link between the user and the serving BS exists, there is a direct association. On the other hand, when the direct link is NLoS and a LoS reflection link exists, the association criterion is an association through RIS reflection. Thus, we have the following lemma.
\begin{lemma}
    Given that the distance between a user and its serving BS is $\xi$, the direct association probability is \vspace{-1mm}
    \begin{equation}
        P_{A_d}^s(\xi)=P_{LoS}(\xi).\vspace{-1mm}
    \end{equation}
\end{lemma}
\begin{lemma} \textit{(Single-Cell Reflection Probability)} \label{reflection probability}
	For a user at a distance $\xi$ from the BS, the probability that there exists at least one RIS reflective LoS link in the cell is 
	\begin{equation}
 \begin{split}
     P_R^{s}(\xi)=&1-\exp(-\\
     &\int_0^{2\pi} {\lambda}^L_R\frac{(\sqrt{R^2-\xi^2\sin^2 \psi}-\xi\cos\psi)^2}{2}\mathrm{d}\psi).
 \end{split}
	\end{equation}
\end{lemma}
\textit{Proof:} See Appendix A.
\begin{lemma}
    Given that the distance between a user and its serving BS is $\xi$, the probability that the serving BS and the user is associated through the reflected LoS link is 
    \begin{equation}
        P_{A_I}^s(\xi)=(1-P_{LoS}(\xi))P_R^s(\xi),
    \end{equation}
where $(1-P_{LoS}(\xi))$ represents the probability that the direct link between the BS and the user is NLoS, and $P_R^s(\xi)$ represents the probability that there exists a reflected LoS link between the user and the NLoS BS.
\end{lemma}

\begin{lemma}
	\textit{(Distance Distribution of Reflective LoS Links)} \label{reflective LoS link dist} Given the distance between BS and a user as $\xi$, the distance between BS and the $l$-th RIS as $\dot{s}_l$, the distance between the user and the $l$-th RIS as $\dot{r}_l$, define $\eta\triangleq\min\limits_l\{\dot{s}_l\cdot \dot{r}_l\}$, its cumulative distribution function is expressed as (\ref{eq:F_eta_xi}), where $\cos(\theta)\in[\frac{s^2+\xi^2}{2s\xi},\frac{s^4+s^2\xi^2-x^2}{2s^3\xi}]$.
\end{lemma}
\textit{Proof:}	See Appendix B.
\begin{figure*}[t]
    \centering
    \begin{equation}\label{eq:F_eta_xi}
		F_{\eta|\xi}(x)=\left\{
		\begin{aligned}
			0 ,& \quad x\leq r\\
			1-\exp(-\int_0^R\int_{\theta}\lambda_RP_{LoS}(\sqrt{s^2+\xi^2-2s\xi\cos\theta})s\mathrm{d}\theta\mathrm{d}s) ,& \quad \text{else}\\
		\end{aligned}
		\right.
	\end{equation}
\end{figure*}
\par
\begin{proposition}
	\textit{(Campbell's Theorem)} \cite{keeler2015campbell}\label{campbell}
	Let $\Phi$ be a point process (PP) in $\mathbb{R}^2$ with density $\lambda(x)$ and $f:\mathbb{R}^2\rightarrow\mathbb{R}$ be a measurable function, then
	\begin{equation}
		\mathbb{E}\big\{\sum\limits_{x_i\in\Phi}f(x_i)\big\}=\int_{\mathbb{R}^2} f(x)\lambda(x)\mathrm{d}x.
	\end{equation}
\end{proposition}
\begin{definition}
	Define the coverage probability as 
	\begin{equation}
		P_{cov}(\gamma_0)=Pr(\gamma>\gamma_0),
	\end{equation}
	where $\gamma_0$ is the setting threshold, and $\gamma$ represents the SINR of the associated link.
\end{definition}
\begin{theorem}
	\textit{(Single-Cell Ergodic Coverage Probability)}\label{eq:coverage single}
	\begin{equation}
		\mathbb{E}[P_{cov}^s(\gamma_0)]=\frac{\int_0^R P_{cov|\xi}^s(\gamma_0)\lambda_u(\xi)2\pi \xi\mathrm{d}\xi}{\int_0^R \lambda_u(\xi)2\pi \xi\mathrm{d}\xi}
	\end{equation}
	where $P_{cov|\xi}^s(\gamma_0)$ is the coverage probability of users at distance $\xi$ from the BS, as expressed in (\ref{eq:P_cov_xi}), and $\tau_1\triangleq\frac{\xi^\beta\sigma^2\gamma_0}{P_0 10^\alpha}$, $\tau_2\triangleq(\frac{P_0 h_s h_r 10^{2\alpha}}{\sigma^2\gamma_0})^\frac{1}{\beta} $,
 \begin{figure*}\vspace{-6mm}
        \centering
        \begin{equation}\label{eq:P_cov_xi}
		\begin{split}
			P_{cov|\xi}^s(\gamma_0)=P_{LoS}(\xi)\exp(-\frac{\tau_1}{N_{BS}N_u})
			+(1-P_{LoS}(\xi))P_R^s(\xi)\int_0^\infty\int_0^\infty F_{\eta|\xi}(\tau_2)f_{h_s}(x_1)\mathrm{d}x_1f_{h_r}(x_2)\mathrm{d}x_2
		\end{split}
	\end{equation}
  {\noindent} \rule[-10pt]{17.5cm}{0.05em}\\
 \end{figure*}
	\begin{equation}
		f_{h_s}(x_1)=\left\{
		\begin{aligned}
			\frac{1}{N_{BS}N_R}\exp(-\frac{1}{N_{BS}N_R} x_1),&\ x_1>0\\
			0,&\ \text{else}
		\end{aligned}
		\right.,
	\end{equation}
	\begin{equation}
		f_{h_r}(x_2)=\left\{
		\begin{aligned}
			\frac{1}{N_RN_{u}}\exp(-\frac{1}{N_RN_{u}} x_2),&\ x_2>0\\
			0,&\ \text{else}
		\end{aligned}
		\right. .
	\end{equation}
\end{theorem}
\textit{Proof:} See Appendix C.
\par
\begin{definition}
    The ergodic rate is defined as 
\begin{equation}
	R=\mathbb{E}[W\log_2(1+\gamma)],
\end{equation}
where $W$ is the bandwidth occupied by each user.
\end{definition} 

\begin{proposition}
	For a positive random variable $X$,\label{positive RV} 
 \begin{equation}
     \mathbb{E}[X]=\int_0^\infty Pr(X>t)\mathrm{d}t .
 \end{equation}
\end{proposition}
\begin{theorem}\textit{(Achievable Rate)}
	The achievable rate of the cell is\label{eq:sumrate single}
	\begin{equation}
		{\Upsilon}=\frac{\int_0^R \int_0^\infty P_{cov|\xi}^s(2^{\frac{t}{W}}-1)\mathrm{d}t \lambda_u(\xi) 2\pi \xi\mathrm{d}\xi}{\int_0^R \lambda_u(\xi)2\pi \xi\mathrm{d}\xi} .
	\end{equation}
\end{theorem}
\textit{Proof:}	See Appendix D.
\vspace{-1mm}
\subsection{Multi-Cell Performance} 
\vspace{2.5mm}
Similarly, we can derive the probability that this typical user has at least one LoS BS and at least one reflective NLoS BS, and further derive the distance distribution of the shortest direct link and the shortest reflective link.
\begin{lemma}
	According to \textbf{Proposition \ref{voidcontact}}, the probability that this typical user has at least one LoS BS equals to 1 minus the void probability,\vspace{-3mm}
	\begin{equation}
		P_L=1-\exp(-\int_0^\infty \lambda_Y^L2\pi \xi\mathrm{d} \xi).
	\end{equation}
\end{lemma}
\begin{lemma}
	Similarly, the probability that this typical user has at least one NLoS BS with the presence of reflective LoS links is\vspace{-1mm}
	\begin{equation}
		P_N^R=1-\exp(-\int_0^\infty \tilde{\lambda}_Y^N 2\pi \xi \mathrm{d}\xi).
	\end{equation}
\end{lemma}
\begin{lemma}
	The distance from the user to the nearest associated LoS BS is $\xi_0=\min\limits_{Y_i\in\Phi_Y^L}\xi_{i}$, and the distribution is the differential of the contact distribution
	\begin{align}
     f_{\xi_0}(x)&=\frac{\mathrm{d}(1-\exp(-\int_0^x \lambda_Y^L2\pi \xi \mathrm{d}\xi))}{\mathrm{d}x}\nonumber\\
     &=\lambda_YP_{LoS}(x)2\pi x \exp(-\int_0^x \lambda_Y^L2\pi \xi\mathrm{d}\xi).
 \end{align}
\end{lemma}
\par
We define the minimum path length product from the user to the reflective NLoS BS as\vspace{-1.5mm}
\begin{equation}
    \eta_0=\min\limits_{Y_j\in\tilde{\Phi}_Y^N}\eta_{j},\ \eta_{j}=\min\limits_{R_l\in\Phi_R^L} r_{l}s_{i,l}.\vspace{-0.5mm}
\end{equation}
It is obtained using a greedy approach: we first find the nearest LoS RIS $R_0\in\Phi_R^L$ to the user, then the distance is denoted as $r_{0}=\min\limits_{R_l\in\Phi_R^L}r_{l} $. Afterwords, we find the nearest BS $Y_0\in\tilde{\Phi}_Y^N $ from $R_0$, and the distance is denoted as $s_0=\min\limits_{Y_i\in\tilde{\Phi}_Y^N}s_{i,0} $, then we have $\eta_0=s_0r_0$.
\begin{lemma}\label{eta0 distribution}
	The distribution of $\eta_0$ is expressed as
 \begin{subequations}
     \begin{equation}
		F_{\eta_0}(x)=\int_0^\infty F_{\eta_0|r_0}(x)f_{r_0}(r)\mathrm{d}r,
	\end{equation} 
	\begin{equation}
		f_{r_0}(x)=\lambda_R P_{LoS}(x)2\pi x\exp(-\int_0^x\lambda_RP_{LoS}(r)2\pi r\mathrm{d}r),
	\end{equation}
	\begin{equation}
		\begin{aligned}
			F_{\eta_0|r_0}(x)&=1-\exp(-\int_{0}^\infty\int_{\arccos(\frac{r_0^2+\xi^2}{2r_0\xi})}^{\arccos(\frac{r_0^4+r_0^2\xi^2-x^2}{2r_0^3\xi})} \tilde{\lambda}_Y^N \xi\mathrm{d}\vartheta\mathrm{d}\xi)\\
            &=1-\exp(-\int_{0}^\infty\int_{0}^{\arccos(\frac{r_0^4+r_0^2\xi^2-x^2}{2r_0^3\xi})} \tilde{\lambda}_Y^N \xi\mathrm{d}\vartheta\mathrm{d}\xi).
		\end{aligned}
	\end{equation}
 \end{subequations}
\end{lemma}
It is worth noting that the distribution represented by the lower case $f$ is the probability density function and the distribution represented by the upper case $F$ is the cumulative density function (CDF).\par
\textit{Proof:} See Appendix E.
\par 
It can be analyzed from the above equation that as the density of the RIS increases, the value interval of the CDF of $\eta_0$ becomes narrower, and $\eta_0$ becomes more concentrated at small values. Further, this leads to smaller path losses in the reflective links, resulting in greater signal power received by the typical user, which demonstrates the benefits of moderately deploying more RISs for communications.
\par
Based on the association criterion in \textbf{Definition \ref{asscocition criterion}}, we now derive the association probability. If the user and the serving BS are directly associated, denoted by $Y_0\in\Phi_Y^L $. That is, the LoS BS has the maximum received signal power, we have\vspace{-1mm}
\begin{equation}
    Y_0=\arg\max\limits_{Y_i\in\Phi_Y^L} g_{d_{i}}\!\!=\arg\max\limits_{Y_i\in\Phi_Y^L}\frac{10^\alpha}{\xi_{i}^\beta}\!=\arg\min\limits_{Y_i\in\Phi_Y^L}\xi_{i}.
\end{equation}

\begin{lemma}\label{eq:P_Ad}
	The probability that the serving BS $Y_0$ is in the LoS BS PPP $\Phi_Y^L $ is \vspace{-1mm}
	\begin{align}
	    P_{A_d}&=P_L(P_N^RPr(\max\limits_{Y_i\in\Phi_Y^L}g_{d_{i}}\!>\!\!\!\max\limits_{Y_j\in\tilde{\Phi}_Y^N}g_{I_{j}}N_R^2)\!+\!1\!-\!P_N^R)\nonumber\\
			&=\int_0^\infty(1-F_{\eta_0}((10^\alpha N_R^2)^\frac{1}{\beta}\xi_0)f_{\xi_0}(\xi_0)\mathrm{d}\xi_0.
	\end{align}
That is, the serving BS is a LoS BS when there is at least one LoS BS ($\Phi_Y^L\neq\varnothing $), and the association criterion is met (i.e., the maximum received signal power from the LoS BSs $\Phi_Y^L$ is greater than that from the reflected NLoS BS $\tilde{\Phi}_Y^N$) or there is no reflected NLoS BS ($\tilde{\Phi}_Y^N=\varnothing $).
\end{lemma}
\par
If the user and the serving BS is associated through a RIS reflective link, denoted by $Y_0\in\tilde{\Phi}_Y^N $, we have\vspace{-1mm}
\begin{equation}
    Y_0=\arg\max\limits_{Y_i\in\tilde{\Phi}_Y^N} g_{I_{i}}\!=\arg\max\limits_{Y_j\in\tilde{\Phi}_Y^N} \frac{10^{2\alpha}}{\eta_{j}^\beta} =\arg\min\limits_{Y_j\in\tilde{\Phi}_Y^N}\eta_{j}.
\end{equation}
\par 
\begin{lemma}\label{eq:P_AI}
	The probability that the serving BS $Y_0$ is in the NLoS BS PPP $\tilde{\Phi}_Y^N$ is \vspace{-1mm}
	\begin{align}
			&P_{A_I}=P_N^R(P_LPr(\max\limits_{Y_i\in\Phi_Y^L}g_{d_{i}}\!<\!\!\max\limits_{Y_j\in\tilde{\Phi}_Y^N}g_{I_{j}}N_R^2)\!+\!1\!-\!P_L)\nonumber\\
			&=\int_0^\infty\!\! F_{\eta_0}((10^\alpha N_R^2)^\frac{1}{\beta}\xi_0)f_{\xi_0}(\xi_0)\mathrm{d}\xi_0\!+\!(1-P_L)P_R^m.
		\end{align}
That is, the serving BS is a NLoS BS when $\tilde{\Phi}_Y^N\neq\varnothing $, and the maximum received signal power from the LoS BSs $\Phi_Y^L$ is less than that from the reflected NLoS BS $\tilde{\Phi}_Y^N$ or $\Phi_Y^L=\varnothing $.
\end{lemma}
\par
Thus, the blind area ratio can be obtained as \vspace{-1mm}
\begin{equation}\label{eq:blindratio}
    P_{blind}=1-P_{A_d}-P_{A_I}.
\end{equation}
Next, we derive the distribution of the distance from the typical user to its serving BS.
\begin{lemma}
	Given that the serving BS $Y_0$ is a LoS BS, we derive the distribution of the distance $\xi_0$ as\vspace{-1mm}
	\begin{align}
			&\tilde{f}_{\xi_0}(x)\nonumber\\
   &=f_{\xi_0}(x)(P_N^RPr(\max\limits_{Y_i\in\Phi_Y^L}g_{d_{i}}>\max\limits_{Y_j\in\tilde{\Phi}_Y^N}g_{I_{j}}N_R^2)\nonumber\\
                &\quad +1-P_N^R)\nonumber\\
                &=f_{\xi_0}(x)Pr(\eta_0>(10^\alpha N_R^2)^\frac{1}{\beta}\xi_0+1-P_N^R)\nonumber\\
			&=f_{\xi_0}(x)(1-F_{\eta_0}((10^\alpha N_R^2)^\frac{1}{\beta}x+1-P_N^R)
		\end{align}
\end{lemma}
\begin{lemma}
	Given that the serving BS $Y_0$ is a NLoS BS, we derive the distribution of $\eta_0$ of the reflective LoS links as\vspace{-1mm}
	\begin{align}
	    &\tilde{f}_{\eta_0}(x)\nonumber\\
   &=\frac{\mathrm{d}F_{\eta_0}(x)}{\mathrm{d}x}(P_LPr(\max\limits_{Y_i\in\Phi_Y^L}g_{d_{i}}<\max\limits_{Y_j\in\tilde{\Phi}_Y^N}g_{I_{j}}N_R^2)\nonumber\\
   &\quad +1-P_L)\nonumber\\
            &=\frac{\mathrm{d}F_{\eta_0}(x)}{\mathrm{d}x}(Pr(\eta_0<(10^\alpha N_R^2)^\frac{1}{\beta}\xi_0)+1-P_L)\nonumber\\
			&=\frac{\mathrm{d}F_{\eta_0}(x)}{\mathrm{d}x} (\int_{(10^\alpha N_R^2)^{-\frac{1}{\beta}}x}^\infty f_{\xi_0}(\xi_0)\mathrm{d}\xi_0+1-P_L)\vspace{-2mm}
	\end{align}
 The proofs of the above two lemmas are similar to those of \textbf{Lemma \ref{eq:P_Ad}} and \textbf{Lemma \ref{eq:P_AI}}, thus we omit the proofs here due to space constraints.
\end{lemma}
\begin{theorem} \textit{(Multi-Cell Ergodic Coverage probability)}\label{eq:coverage multi} The ergodic coverage probability for the typical user is expressed as (\ref{eq:cov_m}).
 \begin{figure*}
     \centering
     \begin{subequations}\label{eq:cov_m}
         \begin{equation}
		\begin{split}
			P_{cov}^m(\gamma_0)&=Pr(\gamma_d>\gamma_0|d)+Pr(\gamma_I>\gamma_0|I)
		\end{split} 
	\end{equation}
\begin{equation}
	\begin{split}
		Pr(\gamma_d>\gamma_0|d)=\int_0^\infty \exp(-\frac{\gamma_0(\sigma^2+Q_1(\xi_0)+Q_2(\xi_0)) \xi_0^\beta}{N_{BS}N_uP_0 10^\alpha}) \tilde{f}_{\xi_0}(\xi_0)\mathrm{d}\xi_0
	\end{split} 
\end{equation}
\begin{equation}
	\begin{split}
		Pr(\gamma_I>\gamma_0|I)=\int_{\frac{\gamma_0(\sigma^2+Q_3(\eta0)+Q_4(\eta_0))\eta_0^\beta}{P_0 10^{2\alpha}}}^\infty  f_{h_{s_{0}}h_{r_{0}}}(z) \mathrm{d}z
	\end{split} 
\end{equation}
\begin{equation}
	\begin{split}
		Q_1(\xi_0)=P_0 \mathbb{E}[h_{d_i}] 10^\alpha \int_{\xi_0}^\infty \frac{1}{\xi^\beta}\lambda_YP_{LoS}(\xi) 2\pi \xi\mathrm{d}\xi\\
	\end{split} 
\end{equation}
\begin{equation}
	\begin{split}
		Q_2(\xi_0)=P_0 \mathbb{E}[h_{s_{j}}]\mathbb{E}[h_{r_{j}}]10^{2\alpha}  \int_{\xi_0}^\infty\int_0^{2\pi}\int_0^\infty \frac{f_R(r)\lambda_Y(1-P_{LoS}(\xi))P_{LoS}(r)}{(r\sqrt{\xi^2+r^2-2\xi r\cos\vartheta})^\beta}\mathrm{d}r \mathrm{d}\vartheta\xi\mathrm{d}\xi
	\end{split} 
\end{equation}
\begin{equation}
	\begin{split}
		Q_3(\eta_0)=10^\alpha P_0 \mathbb{E}[h_{d_i}]\int_{(10^\alpha N_R^2)^{-\frac{1}{\beta}}\eta_0}^\infty\frac{1}{\xi^\beta}\lambda_Y P_{LoS}(\xi) 2\pi \xi\mathrm{d}\xi
	\end{split} 
\end{equation}
\begin{equation}
	\begin{split}
		Q_4(\eta_0)=10^{2\alpha} P_0 \mathbb{E}[h_{s_{j}}]\mathbb{E}[h_{r_{j}}]\int_{(10^\alpha N_R^2)^{-\frac{1}{\beta}}\eta_0}^\infty \int_0^{2\pi}\int_0^\infty \frac{f_R(r)\tilde{\lambda}_Y^N}{(r\sqrt{\xi^2+r^2-2\xi r\cos\vartheta})^\beta}\mathrm{d}r \mathrm{d}\vartheta\xi\mathrm{d}\xi
	\end{split} 
\end{equation}
\begin{equation}
	f_{h_{s_{0}}h_{r_{0}}}(z)=\left\{
	\begin{aligned}
		\frac{1}{N_{BS}N_{R}^2N_u}\int_0^\infty \frac{1}{x} \exp(-\frac{1}{N_{BS}N_{R}} x-\frac{1}{N_RN_{u}}\frac{z}{x})\mathrm{d}x,&\ z>0\\
		0,&\ \text{else}
	\end{aligned} 
	\right.
\end{equation}
\begin{equation}\label{eq:Ev}
	\begin{split}
		\mathbb{E}[h_{v}]=\frac{\psi_t}{2\pi}\frac{\psi_r}{2\pi} M_tM_r+\frac{\psi_t}{2\pi}(1-\frac{\psi_r}{2\pi})M_tm_r+(1-\frac{\psi_t}{2\pi})\frac{\psi_r}{2\pi}m_tM_r+(1-\frac{\psi_t}{2\pi})(1-\frac{\psi_r}{2\pi})m_tm_r 
	\end{split}
\end{equation}
     \end{subequations}
     {\noindent} \rule[-10pt]{17.5cm}{0.05em}
\end{figure*}
The subscripts of (\ref{eq:Ev}) are $[v,t,r]\in\{[d_i,\text{BS},\text{UE}],[s_{j},\text{BS},\text{RIS}],[r_{j},\text{RIS},\text{UE}]\}$.
\end{theorem}
\textit{Proof:} See Appendix F.
\begin{theorem} \textit{(Achievable Rate)} \label{eq:rate_m}
	Using \textbf{Proposition \ref{positive RV}}, the achievable rate of the multi-cell network can be further calculated according to the coverage probability obtained by \textbf{Theorem \ref{eq:coverage multi}},
	\begin{equation}
		\begin{split}
			R&=\mathbb{E}[W\log_2(1+\gamma)]\\&=\int_0^\infty\!\!\!Pr(W\log_2(1+\gamma)>x)dx\\
			&=\int_0^\infty\!\! Pr(\gamma>2^{\frac{x}{W}}-1)dx\\&=\int_0^\infty P_{cov}^m(2^{\frac{x}{W}}-1)dx 
		\end{split}
	\end{equation}
\end{theorem}

\vspace{-1mm}
\section{Simulation Results}\label{sec:simulation_result}
\vspace{3mm}
In this section, we provide numerical results for the ergodic coverage probability and the achievable rate in distributed RISs-assisted mmWave wireless communication systems. We assume that the mmWave communication cellular network works at a frequency of $f_{c}=28$ GHz. The parameters of the large-scale fading channel gain model are set to $\alpha = 5.6$ and $\beta = 2.2$. The average length of the blockage line Boolean model is $\mathbb{E}[L] = 15$ m. Each BS antenna array consists of $N_{BS} = 64$ antennas, while each user is equipped with the number of $N_u = 4$ antennas and is allocated a bandwidth of $W = 200$ MHz.
\vspace{-2mm}
\subsection{Single-Cell Simulation}
\begin{figure}[t]
	\centering
	\vspace{0pt}
	\includegraphics[width=0.5\textwidth]{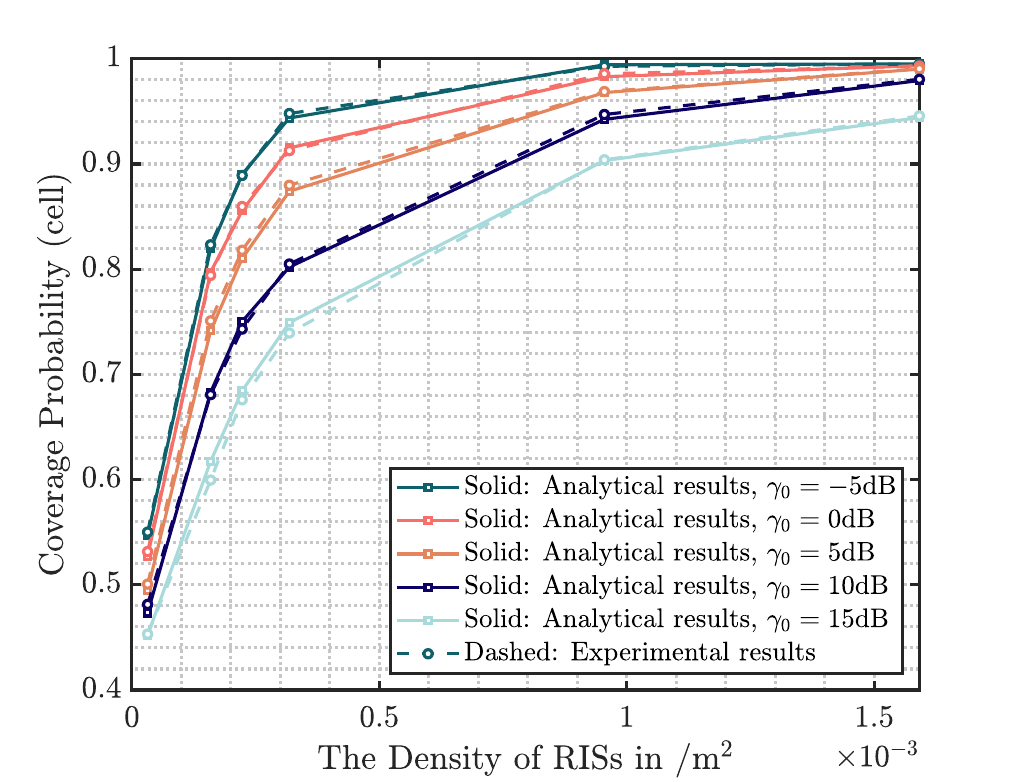}\vspace{-0mm}
	\caption{The impact of the density of RISs on the ergodic coverage probability of the cell.}\label{fig:cov_s}\vspace{-3mm}
\end{figure}

\vspace{2.5mm}
We set the virtual radius as $R = 100$ m and the density of users as $\lambda_u=3.18e$-$3/$m$^2$. Fig.~\ref{fig:cov_s} plots the ergodic coverage probability for different densities of RISs with a blocking density of $\lambda_b=1.59e$-$3/$m$^2$. The solid line represents the ergodic coverage probability in \textbf{Theorem~\ref{eq:coverage single}}, while the corresponding dashed line is obtained by Monte Carlo simulation. The simulation results can overlap the theorem well, which proves the accuracy of our derived formulations. Moreover, we observe that deploying more RISs increases the coverage probability, but the rate of improvement decreases as the density of RISs increases, eventually reaching saturation. Specifically, deploying RISs with a density of $\lambda_R=1.59e$-$4/$m$^2$ (around the number of $5$) can improve the coverage probability by 27.3\%, while deploying RISs with a density of $\lambda_R=9.55e$-$4/$m$^2$ (around $30$) improves the coverage probability by 45.4\%. 
\begin{figure}[t]
	\begin{center}
		\centerline{\includegraphics[width=0.5\textwidth]{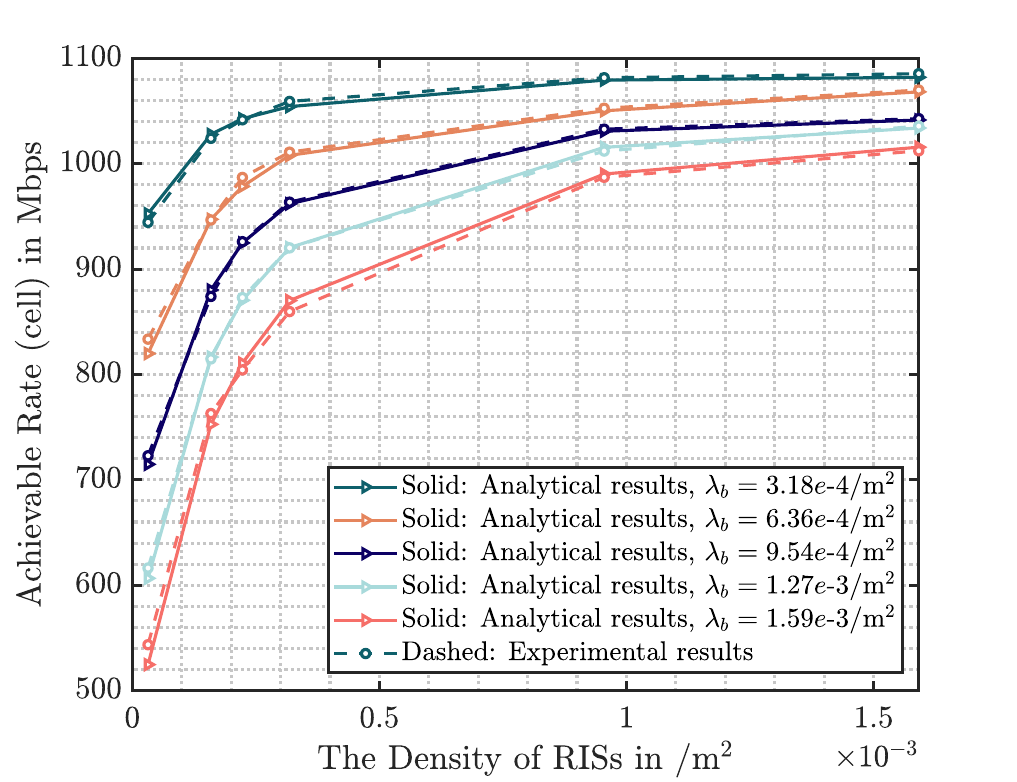}}  \vspace{-0mm}
		\caption{The impact of the density of RISs on the achievable rate for different blockage densities.}
		\label{fig:rate_s} 
	\end{center}\vspace{-6mm}
\end{figure}
\par 
Fig.~\ref{fig:rate_s} plots the effect of the density of deployed RISs on the mean achievable rate for all users in the cell for varying blockage densities. The results of \textbf{Theorem \ref{eq:sumrate single}}, denoted by the solid line, and the Monte Carlo scenario simulation, denoted by the dashed line, are closely matched. As the blockages become denser, the achievable rate decreases, yet the improvement in the achievable rate is more significant with the same density of RISs deployed. For example, when RISs with density $\lambda_R=1.59e$-$3/$m$^2$ (around $50$) are deployed, the achievable rate increases by $129.5$ Mbps (1.14 times) when $\lambda_b = 3.18e$-$4/$m$^2$, and by $491.1$ Mbps (1.93 times) when the blockage density $\lambda_b=1.59e$-$3/$m$^2$. However, it is worth noting that the curve tends to flatten out as the density of RISs increases, which means that the achievable rate does not increase indefinitely as the density of RISs increases, but rather reaches an upper limit.
\vspace{-2.mm}
\subsection{Multi-Cell Network}
\vspace{2.mm}
\par
\begin{figure}[t]
	\begin{center}
		\centerline{\includegraphics[width=0.5\textwidth]{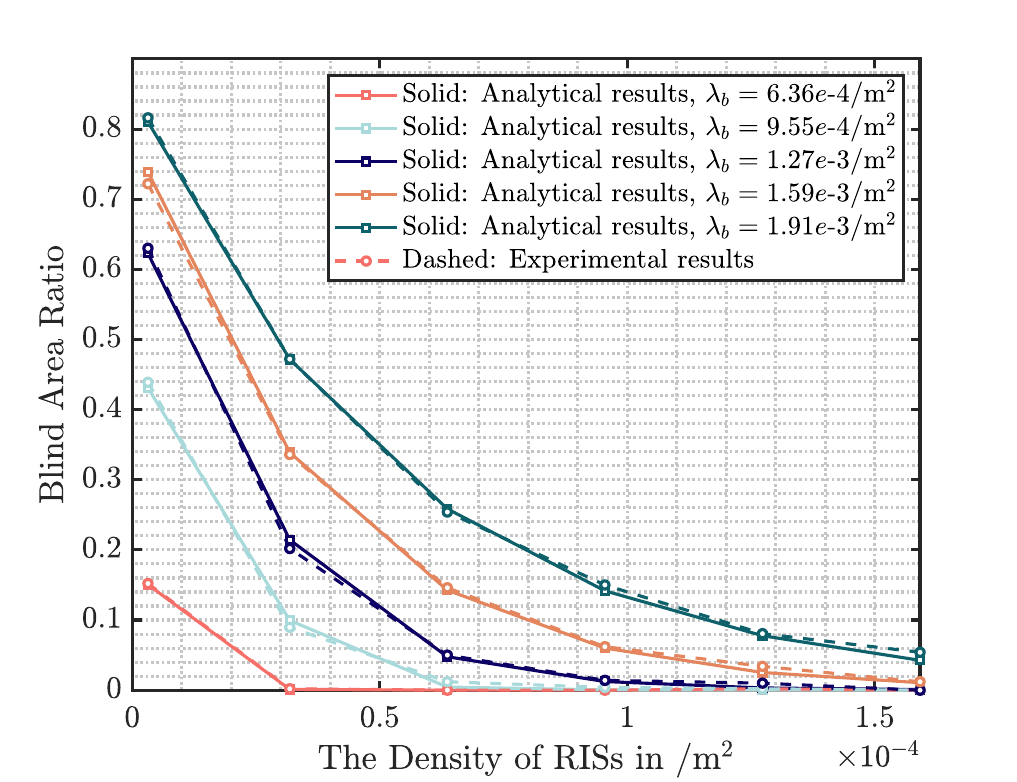}}  \vspace{-0mm}
		\caption{The impact of the RIS deployment on the blind area ratio for different blockage densities.}
		\label{fig:blindration_m} 
	\end{center}\vspace{-5mm}
\end{figure}

Fig.~\ref{fig:blindration_m} plots the effect of the RIS deployment on the blind area ratio for different blockage densities, where the blockage density is set to $1.59e$-$3/$m$^2$. The blind area ratio is the probability that this typical user can not be served by any BS. It can be seen that the reduction in the blind area ratio resulting from the deployment of distributed RISs is more evident at higher blockage densities, which greatly improves the coverage of the cellular network. Specifically, deploying RIS with a density of $\lambda_R = 1.59e$-$4/$m$^2$ reduces the blind area ratio by 15.1\% when the blockage density is $\lambda_b = 6.36e$-$4/$m$^2$, while reduces the blind area ratio by 77.4\% when the blockage density is $\lambda_b = 1.91e$-$3/$m$^2$.
\begin{figure}[t]
	\begin{center}
		\centerline{\includegraphics[width=0.5\textwidth]{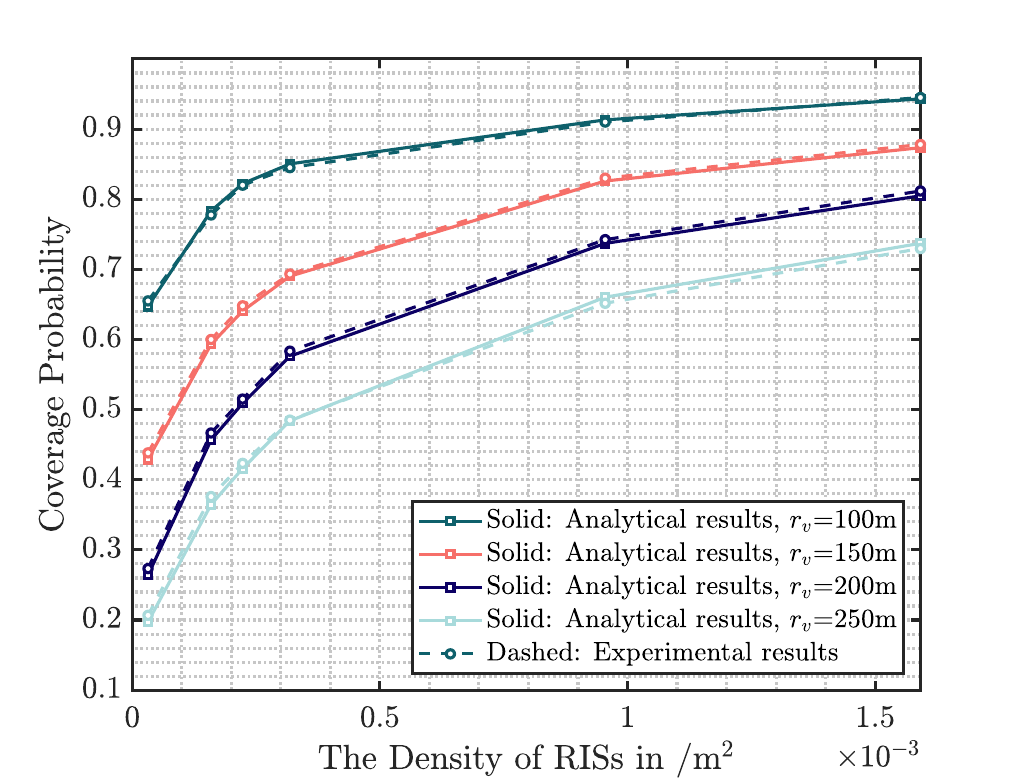}}  \vspace{-0mm}
		\caption{The impact of the RIS deployment on the multi-cell ergodic coverage probability for different BS densities.}
		\label{fig:cov_m_rv} 
	\end{center}\vspace{-6mm}
\end{figure}
\begin{figure}[t]
	\begin{center}
		\centerline{\includegraphics[width=0.5\textwidth]{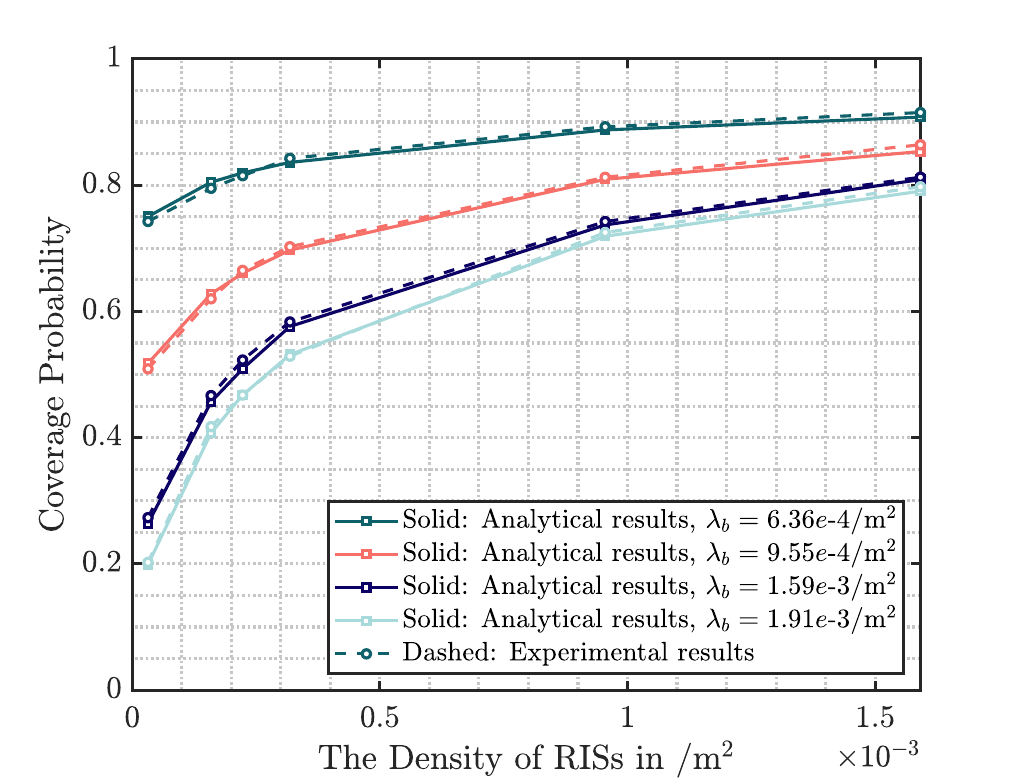}}  \vspace{-0mm}
		\caption{The impact of the RIS deployment on the multi-cell ergodic coverage probability for different blockage densities.}
		\label{fig:cov_m_b} 
	\end{center}\vspace{-5mm}
\end{figure}
\par 
Fig.~\ref{fig:cov_m_rv} and Fig.~\ref{fig:cov_m_b} plot the effect of the RIS deployment density on the ergodic coverage probability for different BS densities and blockage densities, respectively. In Fig.~\ref{fig:cov_m_rv}, we set the blockage density as $\lambda_b = 1.59e$-$3/$m$^2$, and the SINR threshold as $5$ dB. When the BSs density is more sparse, deploying the same density of distributed RISs brings a greater gain in the ergodic coverage probability, which is improved by 27.9\% when $r_v = 100$ m, and by 54.9\% when $r_v = 250$ m. Besides, when the RIS density is $9.55e$-$4/$m$^2$, the ergodic coverage probability is almost maximal, and further increasing the RIS density has little effect on the ergodic coverage probability. Hence, ultra-dense deployment of RISs may not be worthwhile considering the deployment cost. In Fig.~\ref{fig:cov_m_b}, we assume that the virtual cell radius is $r_v = 200$ m, and the SINR threshold is $5$ dB. It can be noticed that when the blockages are sparse, the ergodic coverage probability saturates at a smaller RIS density (less than or equal to $9.55e$-$4/$m$^2$); as the blockages become denser, more RISs are needed for the ergodic coverage probability to reach the upper bound. These values help to reveal how many RISs to deploy in different environments, such as in low-density congested suburban areas and high-density congested urban areas.

\begin{figure}[t]
	\begin{center}
		\centerline{\includegraphics[width=0.5\textwidth]{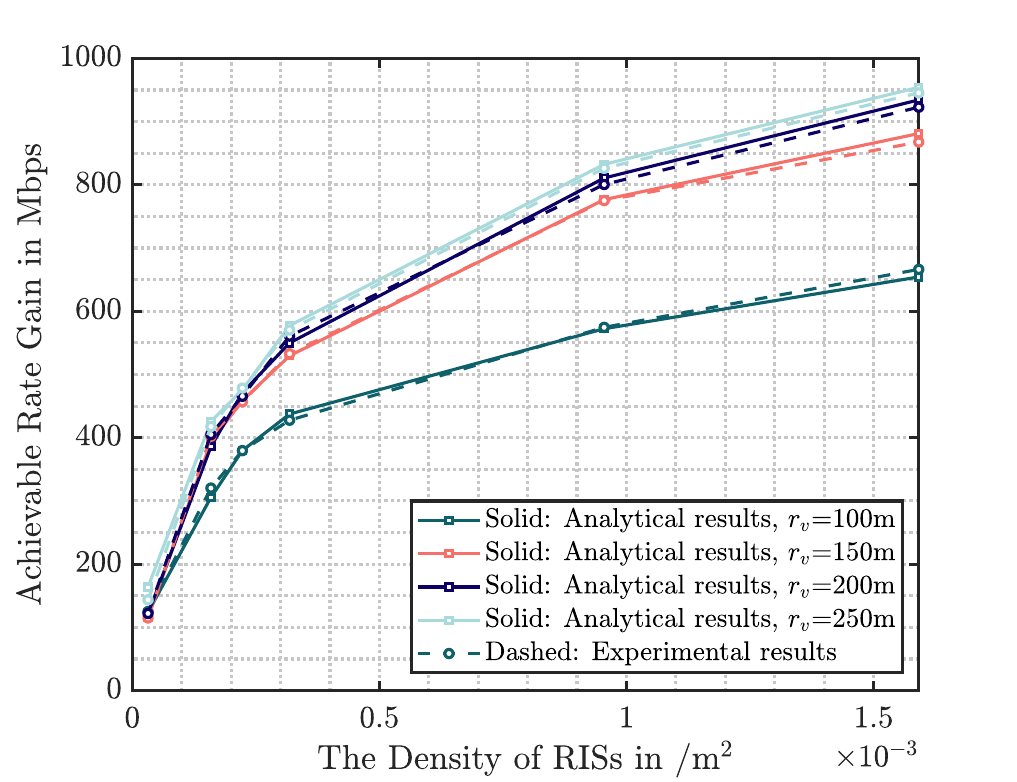}}  \vspace{-0mm}
		\caption{The impact of the RIS deployment on the achievable rate for different BS densities.}
		\label{fig:rate_m_rv} 
	\end{center}\vspace{-7mm}
\end{figure}
\begin{figure}[t]
	\begin{center}
		\centerline{\includegraphics[width=0.5\textwidth]{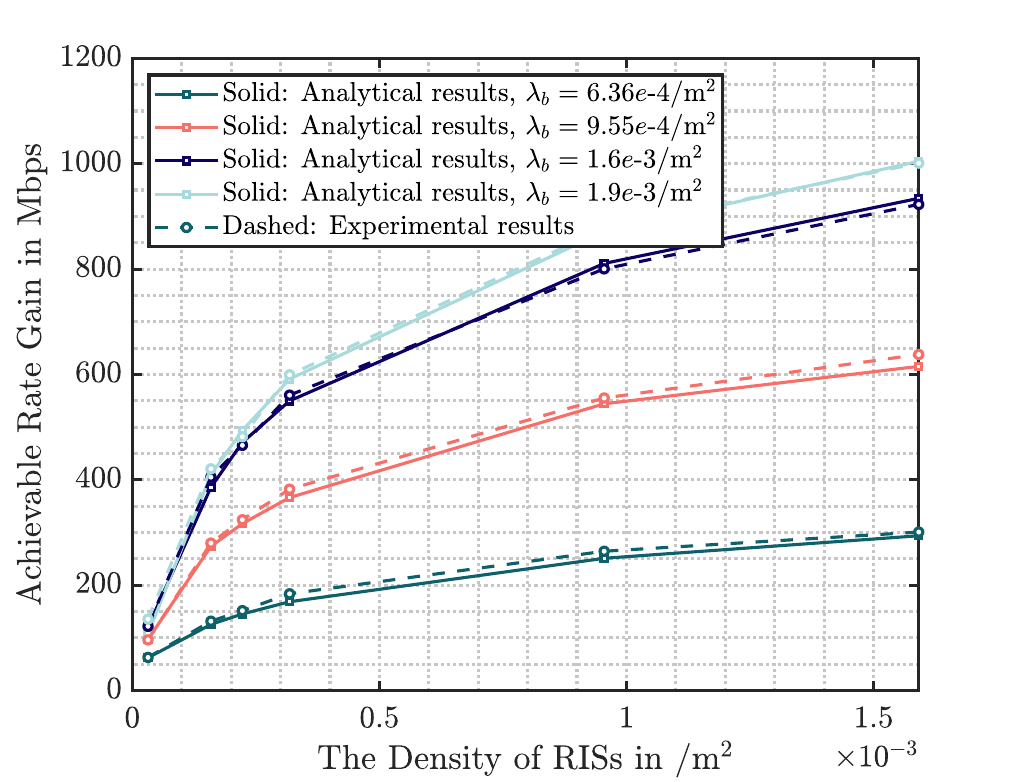}}  \vspace{-0mm}
		\caption{The impact of the RIS deployment on the achievable rate for different blockage densities.}
		\label{fig:rate_m_b} 
	\end{center}\vspace{-6mm}
\end{figure}
\par 
Fig.~\ref{fig:rate_m_rv} and Fig.~\ref{fig:rate_m_b} plot the effect of the RIS deployment density on the achievable rate for different BS densities and different blockage densities, respectively. It can be seen that the growth slope of the multi-cell achievable rate flattens out with the increase of RIS densities. In Fig.~\ref{fig:rate_m_rv}, the blockage density is $\lambda_b = 1.59e$-$3/$m$^2$, 
we can observe that when the RIS density increases from $3.18e$-$5/$m$^2$ to $1.59e$-$3/$m$^2$, the multi-cell achievable rate have an improvement of $527.7$ Mbps ($2.35$ times) and $787.7$ Mbps ($2.82$ times) when the virtual cell radius are $100$ m and $250$ m, respectively. In Fig.~\ref{fig:rate_m_b}, the virtual cell radius is set to be $200$ m, it can be seen that increasing the RIS density from $3.18e$-$5/$m$^2$ to $1.59e$-$3/$m$^2$ can enhance the rate by $199.6$ Mbps ($1.18$ times) and $900.1$ Mbps ($3.29$ times) when $\lambda_b = 6.36e$-$4/$m$^2$  and $\lambda_b = 1.91e$-$3/$m$^2$, respectively.
\vspace{-2.5mm}
\section{Conclusion}\label{sec:conclusion}
\vspace{1.5mm}
In this paper, we leveraged stochastic geometry to investigate the ergodic coverage probability and achievable rate in distributed RISs-assisted single-cell and multi-cell wireless communication scenarios. Firstly, the system model including the distribution models of the blockages, BSs, RISs and users, as well as the association criterion were given. Then, the association probabilities, the distance distributions and the conditional coverage probabilities for two cases, along with the by-product of the blind area ratio, were obtained. Subsequently, the ergodic coverage probabilities and the achievable rate were derived using Campbell's theorem and the total probability theorem.
\par 
The stochastic geometry analysis and simulation results can provide insights into how many distributed RISs should be deployed to achieve a given performance at different blockage densities, which essentially facilitates the optimal design and cost control for practical RISs deployments.

{\appendices
\vspace{-4mm}
\section*{APPENDIX A} 
\section*{Proof of {Lemma \ref{reflection probability}}} Here we calculate the probability that there exists at least one reflective LoS link between the user and RISs within the cell. First, we calculate the number of RISs capable of providing reflective LoS links, which requires a cell-wide integration over the density of LoS RISs ${\lambda}^L_R$. Because ${\lambda}^L_R$ depends on the RIS-user distance $r$, which requires transforming the area into an expression on $r$. To facilitate this transformation, we introduce the angle $\psi=2\pi-\vartheta$, and the shaded area in Fig.~\ref{fig:P_R_s} can be approximated as $\frac{r^2\mathrm{d}\psi}{2} $. Utilizing the Law of Cosines, we can obtain 
\begin{equation}
    \cos(2\pi-\psi)=\cos(\vartheta)=\frac{\xi^2+r^2-R^2}{2\xi r}.
\end{equation}
Further, we derive that $r=\sqrt{R^2-\xi^2\sin^2 \psi}-\xi\cos\psi $. Hence, the number of RISs that can provide reflective LoS links is
\begin{align}
		L^{s}(\xi)=&\int_0^{2\pi} {\lambda}^L_R\frac{r^2\mathrm{d}\psi}{2}\nonumber\\
		=&\int_0^{2\pi} {\lambda}^L_R\frac{(\sqrt{R^2-\xi^2\sin^2 \psi}-\xi\cos\psi)^2}{2}\mathrm{d}\psi.
\end{align}
Thus, the probability is expressed as 
\begin{align}
    &P_R^{s}(\xi)=1-\exp(-L^{s}(\xi))\nonumber\\
		&=1\!-\!\exp(-\!\!\int_0^{2\pi}\!\!\!\! {\lambda}^L_R\frac{(\sqrt{R^2-\xi^2\sin^2 \psi}-\xi\cos\psi)^2\!}{2}\mathrm{d}\psi).
\end{align}

\vspace{-4mm}
\section*{APPENDIX B}
\section*{Proof of {Lemma \ref{reflective LoS link dist}}} As shown in Fig.~\ref{fig:P_R_s}, the angle between the BS-RIS and BS-user links is defined as $\theta$, because the location-dependent thinning is only related to the distance but not to the angle, it can be assumed that $\cos\theta\sim\mathbf{U}[-1,1]$, then we have $\dot{r}_l=\sqrt{\dot{s}_l^2+\xi_l^2-2\dot{s}_l\xi_l\cos\theta}$.
	\begin{align}
		&F_{\eta|\xi}(x)=Pr(\min\limits_l\{\dot{s}_l\cdot \dot{r}_l\}\leq x|\xi)\nonumber\\
            &=1-Pr(\min\limits_l\{\dot{s}_l\cdot \dot{r}_l\}> x|\xi)\nonumber\\
		&\overset{(a)}{=}1-P_{void}(\frac{s^4+s^2\xi^2-x^2}{2s^3\xi} < \cos\theta < \frac{s^2+\xi^2}{2s\xi})\nonumber\\
		&=\left\{
		\begin{aligned}
			0 ,& \quad x\leq \xi\\
			1-\exp(-\int_0^R\int_\theta{\lambda}^L_R s\mathrm{d}\theta \mathrm{d}s) ,& \quad \text{else}\\
		\end{aligned}
		\right.
	\end{align}
The lower bound of equal sign $(a)$ is established because $\min\limits_l\{\dot{s}_l\cdot \dot{r}_l\}>x$. This implies that for $\forall l\in\{1,...,R\}$, there does not exist $\dot{s}_l\cdot \dot{r}_l<x$. Omitting the subscript $l$, $s\cdot r<x$ is equivalent to 
\begin{equation}
    \cos\theta > \frac{s^4+s^2\xi^2-x^2}{2s^3\xi}.
\end{equation}
The upper bound is due to $r^2=s^2+\xi^2-2s\xi\cos\theta>0$. Thus, in the region satisfying
\begin{equation}
    \frac{s^4+s^2\xi^2-x^2}{2s^3\xi} < \cos\theta < \frac{s^2+\xi^2}{2s\xi},
\end{equation}
and in radius $R$, there is no RIS can assist communications. Consequently, the number of points of ${\Phi}^L_R$ is 0.
\vspace{0mm}
\section*{APPENDIX C}
\section*{Proof of {Theorem \ref{eq:coverage single}}} The conditional coverage probability of the direct links is 
\begin{align}
    P_{cov_d|\xi}^s&=Pr(\gamma_d>\gamma_0)=Pr(h_d>\frac{\xi^\beta\sigma^2\gamma_0}{P_0 10^\alpha})\nonumber\\&=1-F_{h_d}(\frac{\xi^\beta\sigma^2\gamma_0}{P_0 10^\alpha}) ,
\end{align}

where
\begin{equation}
	F_{h_d}(x_0)=\left\{
	\begin{aligned}
		1-\exp(-\frac{1}{N_{BS}N_u} x_0),&\ x_0>0\\
		0,&\ \text{else}.
	\end{aligned}\nonumber
	\right.
\end{equation}
\par
When the reflection LoS link with the largest received signal power is selected, the criterion for selecting the associated RIS is
\begin{align}
    l^*=\arg\max\limits_{l\in\{1,...,L\}} \{\frac{10^{2\alpha}}{(\dot{s}_l\dot{r}_l)^\beta}N_R^2\}=\arg\min\limits_{l\in\{1,...,L\}} \{\dot{s}_l\dot{r}_l\},
\end{align}
the SINR of the associated link is expressed as
\begin{align}
    \gamma_I=\frac{10^{2\alpha}P_0  h_{s_{l^*}} h_{r_{l^*}}}{(s_{l^*}r_{l^*})^\beta\sigma^2},
\end{align}
thus the conditional coverage probability of the reflective LoS link is
\begin{align}
    P_{cov_I|\xi}^s=&Pr(\gamma_I>\gamma_0)=Pr(\eta<(\frac{P_0 h_s h_r 10^{2\alpha}}{\sigma^2\gamma_0})^\frac{1}{\beta})\nonumber\\ 
		=&\int_0^\infty\int_0^\infty F_{\eta|\xi}(\tau_2)f_{h_s}(x_1)\mathrm{d}x_1f_{h_r}(x_2)\mathrm{d}x_2.
\end{align}

\par
Therefore, the conditional coverage probability of users at a distance $\xi$ from BS is
\begin{equation}
	P_{cov|\xi}^s(\gamma_0)=P_{A_d}^s(\xi)P_{cov_d|\xi}^s(\gamma_0)+P_{A_I}^s(\xi) P_{cov_I|\xi}^s(\gamma_0).
\end{equation} 
\par
The ergodic coverage probability of the cell is 
\begin{align}
    \mathbb{E}[P_{cov}^s(\gamma_0)]&=\frac{\mathbb{E}[\sum\limits_{u_i\in\Phi_u}P_{cov|\xi}^s(\gamma_0)]}{\int_0^R \lambda_u(\xi)2\pi \xi\mathrm{d}\xi}\nonumber\\
     &\overset{(b)}{=}\frac{\int_0^R P_{cov|\xi}^s(\gamma_0)\lambda_u(\xi)2\pi \xi \mathrm{d}\xi}{\int_0^R \lambda_u(\xi)2\pi \xi\mathrm{d}\xi},
\end{align}

where $(b)$ is based on \textbf{Proposition \ref{campbell}}.
\vspace{-1mm}
\section*{APPENDIX D}
\section*{Proof of {Theorem \ref{eq:sumrate single}}} The random variable $W\log_2(1+\gamma)>0 $, according to \textbf{Proposition \ref{positive RV}}, for a user at a distance $\xi$ from the BS, its ergodic rate is
\begin{align}
    R(\xi)&=\mathbb{E}[W\log_2(1+\gamma)]=\int_0^\infty Pr(W\log_2(1+\gamma)>t)\mathrm{d}t\nonumber\\
		&=\int_0^\infty Pr(\gamma>2^{\frac{t}{W}}-1)\mathrm{d}t\nonumber\\
		&=\int_0^\infty P_{cov|\xi}^s(2^{\frac{t}{W}}-1)\mathrm{d}t .
\end{align}

Since the distance $\xi$ between each user and the BS is also a random variable, the achievable rate of the cell is 
\begin{align}
    {\Upsilon}=&\frac{\sum\limits_{u_i\in\Phi_u} \mathbb{E}[R(\xi)]}{\int_0^R \lambda_u(\xi)2\pi \xi\mathrm{d}\xi}=\frac{\int_0^R R(\xi)\lambda_u(\xi)2\pi \xi\mathrm{d}\xi}{\int_0^R \lambda_u(\xi)2\pi \xi\mathrm{d}\xi}\nonumber\\
		=&\frac{\int_0^R \int_0^\infty P_{cov|\xi}^s(2^{\frac{t}{W}}-1)\mathrm{d}t \lambda_u(\xi)2\pi \xi\mathrm{d}\xi}{\int_0^R \lambda_u(\xi)2\pi \xi\mathrm{d}\xi}.
\end{align}
\vspace{-3.5mm}
\section*{APPENDIX E}
\section*{Proof of {Lemma \ref{eta0 distribution}}} The distribution of the distance of the user to the nearest LoS RIS is
\begin{align}
    f&_{r_0}(x)=\frac{\mathrm{d}(1-\exp(-\int_0^x \lambda_R^L 2\pi r\mathrm{d}r))}{\mathrm{d}x}\nonumber\\
    &=\lambda_R P_{LoS}(x)2\pi x\exp(-\int_0^x\lambda_RP_{LoS}(r)2\pi rdr).
\end{align}

\par
Given a BS is at a distance $\xi$ from the typical user, for any possible realization of $r_0$, we utilize the Law of Cosines and obtain the distance from $R_0$ to the BS as $s=\sqrt{\xi^2+r_0^2-2\xi r_0\cos\vartheta} $. Hence, 
\begin{align}
    &F_{\eta_0|r_0}(x)=Pr(s_0r_0\leq x)=1-Pr(\min\limits_{Y_i\in\tilde{\Phi}_Y^N}s_{i,0}r_0\geq x)\nonumber\nonumber\\
    &\overset{(c)}{=}1-P_{void}(\frac{r_0^4+r_0^2\xi^2-x^2}{2r_0^3\xi}\leq\cos\vartheta\leq\frac{r_0^2+\xi^2}{2r_0\xi})\nonumber\\
    &=1-\exp(-\int_{0}^\infty\int_{\arccos(\frac{r_0^2+\xi^2}{2r_0\xi})}^{\arccos(\frac{r_0^4+r_0^2\xi^2-x^2}{2r_0^3\xi})} \tilde{\lambda}_Y^N \xi\mathrm{d}\vartheta\mathrm{d}\xi)\nonumber\\
        &=1-\exp(-\int_{0}^\infty\int_{0}^{\arccos(\frac{r_0^4+r_0^2\xi^2-x^2}{2r_0^3\xi})} \tilde{\lambda}_Y^N \xi\mathrm{d}\vartheta\mathrm{d}\xi).
\end{align}
\par
We use the void probability in \textbf{Proposition \ref{voidcontact}} in equation $(c)$ because that when the constraints $\frac{r_0^4+r_0^2\xi^2-x^2}{2r_0^3\xi}\leq\cos\vartheta\leq\frac{r_0^2+\xi^2}{2r_0\xi}$ is met, none of $Y_i\in\tilde{\Phi}_Y^N $ satisfies $s_{i,0}r_0\leq x $. Finally, we derive
\begin{align}
    F_{\eta_0}(x)=\int_0^\infty F_{\eta_0|r_0}(x)f_{r_0}(r_0)\mathrm{d}r_0 .
\end{align}
\section*{APPENDIX F}
\section*{Proof of {Theorem \ref{eq:coverage multi}}} We denote the interference signal as $IF_d^L=\sum\limits_{Y_i\in\Phi_Y^L\setminus Y_0} {\frac{10^\alpha}{\xi_i^\beta}P_0 h_{d_i}}$, $IF_d^N=\sum\limits_{Y_j\in\tilde{\Phi}_Y^N} {\frac{10^{2\alpha}}{\eta_j^\beta}P_0 h_{s_{j}}h_{r_{j}}}$, $IF_I^L=\sum\limits_{Y_i\in\Phi_Y^L} {\frac{10^\alpha}{\xi_i^\beta}P_0 h_{d_i}}$, $IF_I^N=\sum\limits_{Y_j\in\tilde{\Phi}_Y^N\setminus Y_0} {\frac{10^{2\alpha}}{\eta_j^\beta}P_0 h_{s_{j}}h_{r_{j}}}$. When the serving BS of the typical user is a LoS BS, the conditional coverage probability is
\begin{align}
    &Pr(\gamma_d>\gamma_0|d)\nonumber\\
  &=\int_0^\infty Pr(h_{d_0}>\frac{\gamma_0(\sigma^2+IF_d^L+IF_d^N)r_0^\beta}{P_0 10^\alpha})\tilde{f}_{\xi_0}(\xi_0)\mathrm{d}\xi_0\nonumber\\
		&=\!\!\int_0^\infty\!\!\!\! \mathbb{E}_{\Phi_Y}[\exp(-\frac{\gamma_0(\sigma^2\!+\!IF_d^L\!+\!IF_d^N)\xi_0^\beta}{N_{BS}N_uP_0 10^\alpha})]\tilde{f}_{\xi_0}(\xi_0)\mathrm{d}\xi_0.
\end{align}

We have
\begin{align}
    &\mathbb{E}_{\Phi_Y}[\exp(-\frac{\gamma_0(\sigma^2+IF_d^L+IF_d^N)\xi_0^\beta}{N_{BS}N_uP_0 10^\alpha})]\nonumber\\
		=& \exp(-\frac{\gamma_0(\sigma^2+\mathbb{E}_{\Phi_Y^L}[IF_d^L]+\mathbb{E}_{\tilde{\Phi}_Y^N}[IF_d^N]) \xi_0^\beta}{N_{BS}N_uP_0 10^\alpha}).
\end{align}
Since the large-scale fading channel gain and the small-scale fading channel gain are independent of each other, according to \textbf{Proposition \ref{campbell}}, we derive
    \begin{equation}\label{eq:Q1}
	\begin{split}
		&\mathbb{E}_{\Phi_Y^L}[IF_d^L]\\
  &=\mathbb{E}_{\Phi_Y^L}[\sum\limits_{Y_i\in\Phi_Y^L\setminus Y_0} \frac{10^\alpha}{\xi_i^\beta}P_0 h_{d_i}]\\
		&= P_0 \mathbb{E}[h_{d_i}] 10^\alpha \int_{\xi_0}^\infty \frac{1}{\xi^\beta}\lambda_Y^L 2\pi \xi\mathrm{d}\xi\\
            &= P_0 \mathbb{E}[h_{d_i}] 10^\alpha \int_{\xi_0}^\infty \frac{1}{\xi^\beta}\lambda_YP_{LoS}(\xi) 2\pi \xi\mathrm{d}\xi\\
		&\triangleq Q_1(\xi_0),
	\end{split} 
\end{equation}
\begin{equation}\label{eq:Q2}
	\begin{split}
		&\mathbb{E}_{\tilde{\Phi}_Y^N}[IF_d^N]\\
  &=\mathbb{E}_{\tilde{\Phi}_Y^N}[\sum\limits_{Y_j\in\tilde{\Phi}_Y^N} \frac{10^{2\alpha}}{\eta_j^\beta}P_0 h_{s_{j}}h_{r_{j}}]\\
		&=P_0 \mathbb{E}[h_{s_{j}}]\mathbb{E}[h_{r_{j}}]10^{2\alpha} \int_{\xi_0}^\infty \frac{1}{(rs)^\beta}\tilde{\lambda}_Y^N 2\pi \xi\mathrm{d}\xi\\
		&=P_0 \mathbb{E}[h_{s_{j}}]\mathbb{E}[h_{r_{j}}]10^{2\alpha}\\
  &\ \ \times \int_{\xi_0}^\infty\!\!\!\! \int_0^{2\pi}\!\!\!\!\int_0^\infty \!\!\frac{f_R(r)\tilde{\lambda}_Y^N}{(r\sqrt{\xi^2+r^2-2\xi r\cos\vartheta})^\beta}\mathrm{d}r \mathrm{d}\vartheta\mathrm{d}\xi\\ 
		&=P_0 \mathbb{E}[h_{s_{j}}]\mathbb{E}[h_{r_{j}}]10^{2\alpha}\\&\ \ \times  \int_{\xi_0}^\infty\!\!\!\!\int_0^{2\pi}\!\!\!\!\int_0^\infty \!\!\frac{f_R(r)\lambda_Y(1\!-\!P_{LoS}(\xi))P_{LoS}(r)\!\!}{(r\sqrt{\xi^2+r^2-2\xi r\cos\vartheta})^\beta}\mathrm{d}r \mathrm{d}\vartheta\mathrm{d}\xi\\&\triangleq Q_2(\xi_0),
	\end{split} 
\end{equation}
where $f_R(r)$ is the distance distribution from points in the PPP $\Phi_R^L$ to the origin whose square satisfies the uniform distribution.
\par
When the serving BS of the typical user is a NLoS BS, the conditional coverage probability is expressed as
	\begin{align}
	    &Pr(\gamma_I>\gamma_0|I)\nonumber\\
  &=\int_0^\infty \!\!Pr(h_{s_{0}}h_{r_{0}}>\frac{\gamma_0(\sigma^2+IF_I^L+IF_I^N)\eta_0^\beta}{P_0 10^{2\alpha}})\tilde{f}_{\eta_0}(\eta_0)\mathrm{d}\eta_0\nonumber\\&
        =\int_{\mathbb{E}_{\Phi_Y}[\frac{\gamma_0(\sigma^2+IF_I^L+IF_I^N)\eta_0^\beta}{P_0 10^{2\alpha}}]}^\infty  f_{h_{s_{0}}h_{r_{0}}}(z) \mathrm{d}z\nonumber\\
        &=\int_{\frac{\gamma_0(\sigma^2+\mathbb{E}_{\Phi_Y^L}[IF_d^L]+\mathbb{E}_{\tilde{\Phi}_Y^N}[IF_d^N])\eta_0^\beta}{P_0 10^{2\alpha}}}^\infty  f_{h_{s_{0}}h_{r_{0}}}(z) \mathrm{d}z,
	\end{align}
\begin{equation}
	\begin{split}
		\text{with}\quad &\mathbb{E}_{\Phi_Y^L}[IF_d^L]\\
  &=\mathbb{E}_{\Phi_Y^L}[\sum\limits_{Y_i\in\Phi_Y^L} \frac{10^\alpha}{\xi_i^\beta}P_0 h_{d_i}]\\&
		=10^\alpha \int_{(10^\alpha N_R^2)^{-\frac{1}{\beta}}\eta_0}^\infty \frac{1}{\xi^\beta}\lambda_Y^L 2\pi \xi\mathrm{d}\xi P_0 \mathbb{E}[h_{d_i}]\\
		&=10^\alpha P_0 \mathbb{E}[h_{d_i}]\int_{(10^\alpha N_R^2)^{-\frac{1}{\beta}}\eta_0}^\infty\frac{1}{\xi^\beta}\lambda_Y P_{LoS}(\xi) 2\pi \xi\mathrm{d}\xi\\&\triangleq Q_3(\eta_0),
	\end{split} 
\end{equation}
\begin{equation}
	\begin{split}
		\text{and}\quad &\mathbb{E}_{\tilde{\Phi}_Y^N}[IF_d^N]\\
  &=\mathbb{E}_{\tilde{\Phi}_Y^N}[\sum\limits_{Y_j\in\tilde{\Phi}_Y^N\setminus Y_0} \frac{10^{2\alpha}}{\eta_j^\beta}P_0 h_{s_{j}}h_{r_{j}}]\\&
		=10^{2\alpha} \int_{(10^\alpha N_R^2)^{-\frac{1}{\beta}}\eta_0}^\infty \frac{1}{\eta^\beta}\tilde{\lambda}_Y^N 2\pi \xi\mathrm{d}\xi P_0 \mathbb{E}[h_{s_{j}}]\mathbb{E}[h_{r_{j}}] \\
		&=10^{2\alpha} P_0 \mathbb{E}[h_{s_{j}}]\mathbb{E}[h_{r_{j}}]\int_{(10^\alpha N_R^2)^{-\frac{1}{\beta}}\eta_0}^\infty \int_0^{2\pi}\int_0^\infty\\&\quad\  \frac{f_R(r)\tilde{\lambda}_Y^N}{(r\sqrt{\xi^2+r^2-2\xi r\cos\vartheta})^\beta}\mathrm{d}r \mathrm{d}\vartheta\xi\mathrm{d}\xi\\&\triangleq Q_4(\eta_0).
	\end{split} 
\end{equation}

According to the \textbf{Assumption \ref{smallscale}},
\begin{align}
    &\mathbb{E}[h_{v}]=\frac{\psi_t}{2\pi}\frac{\psi_r}{2\pi} \mathbb{E}[\exp(\frac{1}{M_tM_r\!})]\!+\!\frac{\psi_t}{2\pi}(1\!-\!\frac{\psi_r}{2\pi})\mathbb{E}[\exp(\frac{1}{M_tm_r\!})]\nonumber\\
		&\!+\!(\!1\!-\!\frac{\psi_t}{2\pi}\!)\frac{\psi_r}{2\pi}\mathbb{E}[\exp(\frac{1}{m_tM_r\!\!})]\!+\!(\!1\!-\!\frac{\psi_t}{2\pi}\!)(\!1\!-\!\frac{\psi_r}{2\pi}\!)\mathbb{E}[\exp(\frac{1}{m_tm_r\!\!})]\nonumber\\
		&=\frac{\psi_t}{2\pi}\frac{\psi_r}{2\pi} M_tM_r+\frac{\psi_t}{2\pi}(1-\frac{\psi_r}{2\pi})M_tm_r\nonumber\\
		&+(1-\frac{\psi_t}{2\pi})\frac{\psi_r}{2\pi}m_tM_r+(1-\frac{\psi_t}{2\pi})(1-\frac{\psi_r}{2\pi})m_tm_r,
\end{align}
where $[v,t,r]\in\{[d_i,\text{BS},\text{UE}],[s_{j},\text{BS},\text{RIS}],[r_{j},\text{RIS},\text{UE}]\}$. Since
\begin{equation}\nonumber
	f_{h_{s_{0}}}(x)=\left\{
	\begin{aligned}
		\frac{1}{N_{BS}N_{R}}\exp(-\frac{1}{N_{BS}N_{R}} x),&\ x>0\\
		0,&\ \text{else}
	\end{aligned} 
	\right.,
\end{equation}
\begin{equation}\nonumber
	f_{h_{r_{0}}}(x)=\left\{
	\begin{aligned}
		\frac{1}{N_RN_{u}}\exp(-\frac{1}{N_RN_{u}}x),&\ x>0\\
		0,&\ \text{else}
	\end{aligned} 
	\right.,
\end{equation}
we have
\begin{align}
    &f_{h_{s_{0}}h_{r_{0}}}(z)=\int_0^\infty \frac{1}{|x|}f_{h_{s_{0}}}(x)f_{h_{r_{0}}}(\frac{z}{x})\mathrm{d}x\nonumber\\
        &\!=\!\frac{1}{N_{BS}N_{R}^2N_u}\!\!\int_0^\infty \!\!\frac{1}{x} \exp(-\frac{1}{N_{BS}N_{R}} x\!-\!\frac{1}{N_RN_{u}}\frac{z}{x})\mathrm{d}x.
\end{align}
Therefore, we derive the coverage probability equation as
\begin{equation}
	\begin{split}
		P_{cov}^m(\gamma_0)&=Pr(\gamma_d>\gamma_0|d)+Pr(\gamma_I>\gamma_0|I).
	\end{split} 
\end{equation}
}

\bibliography{IEEEabrv,myrefs}
\bibliographystyle{IEEEtran}
\end{document}